\documentclass[11pt]{article}
\usepackage[utf8]{inputenc}

\usepackage[paper=a4paper,margin=1in]{geometry}
\usepackage[header,title,page,titletoc]{appendix} 

\usepackage{amsmath}
\usepackage{amsfonts}
\usepackage{float}
\usepackage{braket}
\usepackage{graphicx}
\usepackage{subcaption}
\usepackage[none]{hyphenat} 

\newcommand\eea{\end{eqnarray}}
\newcommand\bea{\begin{eqnarray}}
\numberwithin{equation}{section}
\newcommand\nn{\nonumber}
\newcommand{\be}{\begin{equation}}
\newcommand{\ee}{\end{equation}}
\newcommand{\tr}{\text{tr} }
\def\({\left(}
\def\){\right)}
\def\[{\left[}
\def\]{\right]}

\usepackage[dvipsnames]{xcolor}
\definecolor{rossocorsa}{rgb}{0.83, 0.0, 0.0}
\definecolor{navyblue}{rgb}{0.0, 0.0, 0.5}

\usepackage{hyperref}
\hypersetup{
    colorlinks,
    citecolor=rossocorsa,
    filecolor=navyblue,
    linkcolor=navyblue,
    urlcolor=navyblue
}

\begin{document} 

\begin{titlepage}

\begin{center}

\phantom{ }

{\bf \LARGE{Rényi entropies in the $n\to0$ limit and\\ \vspace{.2cm} entanglement temperatures}}

\vskip 1cm

Cesar A. Agón${}^{\dagger}$,  Horacio Casini${}^{*}$, Pedro J. Martinez${}^{\ddagger}$

\vskip 1cm

\small{  \textit{Instituto Balseiro, Centro At\'omico Bariloche}}

\vskip .15cm

\small{\textit{ 8400-S.C. de Bariloche, R\'io Negro, Argentina}}

\vskip 3cm

\begin{abstract}
Entanglement temperatures (ET) are a generalization of Unruh temperatures valid for states reduced to any region of space. They encode in a thermal fashion the high energy behavior of the state around a point. These temperatures are determined by an eikonal equation in Euclidean space. We show that the real-time continuation of these equations implies ballistic propagation. For theories with a free UV fixed point, the ET determines the state at a large modular temperature. In particular, we show that the $n \to 0$ limit of Rényi entropies $S_n$, can be computed from the ET. This establishes a formula for these Rényi entropies for any region in terms of solutions of the eikonal equations. In the $n\to 0$ limit, the relevant high-temperature state propagation is determined by a free relativistic Boltzmann equation, with an infinite tower of conserved currents. For the special case of states and regions with a conformal Killing symmetry, these equations coincide with the ones of a perfect fluid. 
\end{abstract}
\end{center}

\vspace{2cm}

\small{\vspace{7 cm}\noindent ${}^{\dagger}$cesar.agon@cab.cnea.gov.ar\\\noindent ${}^{*}$casini@cab.cnea.gov.ar\\
${ }^{\ddagger}$pedro.martinez@cab.cnea.gov.ar
}

\end{titlepage}

\setcounter{tocdepth}{2}

{\parskip = .4\baselineskip \tableofcontents}
\newpage


\section{Introduction}
\label{Sec-uno}

In quantum field theory (QFT) there is a natural notion of subsystems labeled by regions $V$  of the space. There is an algebra of operators corresponding to any of these regions. Understanding the properties of the vacuum state reduced to these algebras has been of much recent interest. In particular, much attention has been given to statistical measures such as the entanglement entropy and Rényi entropies. One feature that is general to all QFTs is that these reduced states display local thermal-like properties at high energies. To be more precise, if in the subsystem $V$ we compute the relative entropy between the vacuum $\omega$ and another state $\sigma$ given by a high energy localized excitation around a point $x$, we get a linear dependence with the energy $E$ of the excitation
\be
S(\sigma|\omega)\sim \beta(x,\hat{p}) \,E\,, \label{hht}
\ee
where $\hat{p}$ is the direction of the excitation momentum \cite{Arias:2016nip}. 
This follows from the monotonicity of relative entropy and the known form of the vacuum state for a Rindler wedge (where $V$ is one-half of the space divided by a plane). The coefficient $\beta(x,\hat{p})$ can be thought of as a position and direction-dependent inverse temperature determining the Boltzmann damping factor of the state at high energies around $x$. We will call $\beta(x,\hat{p})^{-1}$ entanglement temperatures (ET) because they are produced by entanglement with the complement of $V$. For free fields $\beta(x,\hat{p})$ gives the structure of the local terms in the modular Hamiltonian.   

Eq. (\ref{hht}) is a generalization of Unruh temperatures that hold for the Rindler wedge, and of Hawking temperature for an evaporating black hole. Temperature is a concept associated with equilibrium. For a system that is out of equilibrium temperature can only be an approximate notion. For a radiating hot body, for example, black body factors will not allow the extraction of a temperature from the spectrum at low energies, and at arbitrarily large energies we do not expect to get a precise Boltzmann damping. However, if the space is non-dynamical, the rigidity of space localization produces such a precise notion of high-energy thermal behavior.  This is clear from Hawking's calculation of the radiation from a black hole evaporating in the vacuum. The same example shows the necessity that the ET should be direction-dependent since in the asymptotic region such temperatures can only be nonzero in the direction of the black hole.      

For the case of free fields, or, more generally theories with a free UV fixed point, and a static geometry, the ET were shown to be universal. Their value follows from the geometric data from the solutions of a system of non-linear partial differential equations in Euclidean space (eikonal equations) \cite{arias2017anisotropic}. In this paper we first make further progress in understanding the structure of these equations, showing that the ET are invariant under certain canonical transformations of the eikonal equations. Then, we compute how the ET propagates in real-time and in non-static situations. This allows us to compute them in the black hole evaporation scenario from simple geometric considerations.      

 The ET then gives a glimpse at the high-energy behavior of the reduced density matrices. A different way to access this behavior is provided by the Rényi entropies. These are defined as 
 \be
S_n=\frac{1}{1-n}\, \log \tr \rho^n\,.
  \ee
If we write $\rho= c\, e^{-K}$, with $K$ the modular Hamiltonian, we see the index $n$ acts as inverse modular temperature for a state proportional to $\rho^n$. The limit $n\to 0$ corresponds to the limit of large modular temperature. For theories with a free UV fix point, we compute the $n\rightarrow 0$ limit of the Rényi entropy in terms of the ET. This gives these Rényi entropies in terms of a solution to a problem on partial differential equations. We check this relation for the cases where both the ET and the Rényi entropies are known independently.  
  
In contrast to the entanglement entropy, which measures correlations in the vacuum, the $n\to 0$ limit of the Rényi entropy, which for simplicity we will call Rényi-$0$, is a quantity related to a highly excited state. For UV-free theories, this state can be understood in terms of the ET. It obeys a relativistic free Boltzmann equation. Besides having a conserved entropy current and stress tensor, it has an infinite tower of further conserved higher spin currents. For the cases where there is a conformal symmetry of the reduced state this excited state can be described by the equations of a perfect fluid, but this is not true for other cases. In the Euclidean formulation, the description is in terms of a stationary fluid with vortex-like boundary conditions.  We end by briefly discussing the expectations for the corresponding behavior of non-free conformal field theories (CFT) in this $n\to 0$ limit.

\section{Structure of the density matrix at high energies}\label{Sec2}

For a free field, the vacuum is a Gaussian state. In consequence, the modular Hamiltonian $K$ corresponding to a spatial region $V$ is quadratic in the field. This quadratic expression can be diagonalized by a Bogoliubov transformation,
\be
K= \int d\lambda\int ds\,\, (2\pi s) \, a_{s,\lambda}^\dagger a_{s,\lambda}\,,\label{dsds}
\ee
giving to the density matrix $\rho\sim e^{-K}$ the form of a decoupled set of thermalized oscillators.
The parameter $s$ gives the ``modular energy'' of the modes while $\lambda$ are additional degeneracies. The mode operators $a_{s,\lambda}$ are linear combinations of the fields in $V$. The expression for these modes follows by solving the free field equation in Euclidean space where multiplicative boundary conditions are imposed on the region $V$ at $x^0=0$. See \cite{arias2017anisotropic} for details. 
 For example, for a massless scalar, we have to solve
\begin{equation}
\square \phi =0 \,,\qquad\qquad \phi^+(\vec{x}) = e^{-2 \pi s}\, \phi^{-}(\vec{x})\,, \qquad\qquad x^0=0\,, \qquad \vec{x}\in V\,. \label{ddd}
\end{equation}
Here $\phi^\pm$ are the boundary values of the field as we approach $V$ from positive and negative $x^0$. These same boundary values are used to construct the eigenmodes of $K$.

\subsection{Euclidean eikonal equations}

If we are interested in the large eigenvalue limit $|s| \gg 1$ we can use an eikonal approximation to (\ref{ddd}). Writing 
\be
\phi(x)=f(x)\,e^{s \,\alpha(x)}\,,\label{177}
\ee
where $f(x)$ is a slowly varying function, we get in the large $|s|$ limit the Euclidean eikonal equation
\be
(\nabla \alpha)^2=0\,. \label{df}
\ee
$\alpha$ is a complex-valued function having a cut at  $x^0=0,\, \vec{x}\in V$, where boundary conditions 
 \be
 \alpha^+=\alpha^--2\pi \label{bc}
 \ee
 are imposed on the limit values $\alpha^\pm$ of $\alpha$ on the two sides of the cut.
It is useful to separate the real and imaginary parts 
\be
\alpha=a+i \, b\,,
\ee 
and define 
\be\label{ABdef}
A=\nabla a\,, \qquad B=\nabla b\,.
\ee
With this, we can alternatively write the eikonal equation as one of two closed real one-forms in $\mathbb{R}^d-\partial V$ such that   
\begin{equation}
   d\wedge A=0 \qquad d\wedge B=0\,, 
   \qquad A^2=B^2\,,\qquad A\cdot B=0\,, \label{uno}
\end{equation}
\begin{equation}
   \oint_\Gamma A\cdot dx=2 \pi\,,\qquad \oint_\Gamma B\cdot dx=0\,,\label{dos}
\end{equation}
where $\Gamma$ is any curve simply linked with $\partial V$. Equation \eqref{dos} is fixed so that the circulation of $A$ around $\Gamma$ going through $V$ in the positive time direction is $2\pi$.

The orthogonal vectors $A, B$ are singular on $\partial V$, where they have a vortex-like singularity, but they are non-singular elsewhere. 
By symmetry, it follows that on the plane $x^0=0$ we can take 
\be
A(x)=\{A^0(\vec{x}),0\}\,,\qquad A^0(\vec{x})>0\,,\qquad  B(x)=\{0,\vec{B}(\vec{x})\}\,,\qquad \vec{x}\in V\,.\label{ot}
\ee
These eikonal equations for the large eigenvalues of the modular Hamiltonian are the same for all free fields independently of the spin and mass. The mass is irrelevant in this limit of high gradient solutions of the form (\ref{177}).    

\subsection{Entanglement temperatures and high energy density matrix}
 
For each $\vec{x}\in V$ and each direction $\hat{p}$ at this point, there is a solution of the eikonal equations such that $\hat{B}(\vec{x})=\hat{p}$. Using this, we can define an {\sl entanglement temperature}\footnote{The entanglement temperatures were previously called null temperatures or local temperatures \cite{Arias:2016nip,arias2017anisotropic}. We prefer the present name since it refers more properly to their origin from entanglement.}  $T(\vec{x},\hat{p})$, that depends on the point $\vec{x}\in V$ and direction $\hat{p}$, as
\be
T(\vec{x}, \hat{p})=\frac{|A(\vec{x})|}{2\pi}=\frac{|B(\vec{x})|}{2 \pi}\,,\qquad \hat{p}=\hat{B}(\vec{x}) \,,\qquad\vec{x}\in V\,.\label{14}
\ee
We will call
\be
\beta(\vec{x},\hat{p})=T(\vec{x}, \hat{p})^{-1}
\ee
to the inverse temperature.

The direction $\hat{p}$ selects the particular eikonal solution we have to use in (\ref{14}). Coming back to the expression of the density matrix (\ref{dsds}), it follows that the ``high energy'' contribution of the large eigenvalues is of the form \cite{arias2017anisotropic}
\be
K_{HE}=\int_V d^{d-1}x\, \int d^{d-1}p\, \,\beta(\vec{x},\hat{p}) \,\, |\vec{p}|\,   a_{\vec{p},\vec{x}}^\dagger \, a_{\vec{p},\vec{x}}\,. \label{esaa}
\ee
Here the creation and annihilation operators are the ones coming from the decomposition of the field into localized high-energy wave packets. The integrals in phase space are to be thought as discrete sums where there are $(2\pi)^{-(d-1)} \,d^{d-1}x\, d^{d-1}p$ independent, canonically normalized modes per volume element of phase space. The same expression holds for fermions and bosons, and spin labels have to be added for fields with spin. The expression (\ref{esaa}) gives a high energy sector of the reduced density matrix that is locally a thermal state with a direction-dependent inverse temperature $\beta(\vec{x},\hat{p})$. This temperature follows directly from the solutions of the eikonal equations.

Eq. (\ref{esaa}) generalizes Unruh temperature for regions of arbitrary shapes provided we look at high energy localized excitations. These are angle-dependent Unruh temperatures. To make this idea precise, we can use the relative entropy. Consider a high energy localized excitation at a point $\vec{x}\in V$, with momentum $p=\{p^0,\vec p\}$, with $p^0\sim |\vec{p}|$, in the limit of large $p^0$. The relative entropy $S_{\rm rel}=S(\sigma|\omega)$ in $V$ between the excited state $\sigma$ and the vacuum state $\omega$ is dominated by the expectation value of the modular Hamiltonian and results \cite{Arias:2016nip}
\be
S_{\rm rel}\sim \beta(\vec{x},\hat{p}) \, p^0\,.\label{eqq}
\ee
This is the same as the relative entropy of the excitation with a truly thermal state with the same $\beta$. Eq. (\ref{eqq}) gives a precise meaning to the entanglement temperatures. 

Though it is not evident from the eikonal equations, it follows from monotonicity of relative entropy that $\beta(\vec{x},\hat{p})$, for fixed $\vec{x},\hat{p}$, increases when changing the region $V$ to a larger one $W$, $V\subseteq W$. The temperatures, produced by entanglement with the complementary region,  decrease as the boundaries of the regions get farther away.  

\subsection{Canonical transformations}

The Euclidean eikonal equation $(\nabla \alpha)^2=0$ for complex $\alpha$ is a cousin of the corresponding eikonal equation for real time. This later is a particular case of the Hamilton-Jacobi equation, namely, the one for free relativistic particles. We should have a complete set of solutions $\alpha(x,k)$, where $k$ are parameters. In the present problem, the set $k$ has $d-2$ dimensions, allowing a solution for each direction $\hat{B}(\vec{x})=\hat{p}$ for a fixed point $\vec{x}\in V$.  As in the Hamilton-Jacobi equation, these solutions can be used to produce more solutions by ``canonical transformations'' in the following way, see e.g. \cite{frittelli1999eikonal}. We write
\be
\tilde{\alpha}(x,k)=\alpha(x,k)+ f(k)\,,
\ee
for an arbitrary function $f$. Then we demand 
\be
\nabla_k \, \tilde{\alpha}(x,k)=0\,.
\ee
Solving algebraically for $k(x)$ from this equation, the new solutions of the eikonal equation are  
\be
\tilde{\tilde{\alpha}}(x)=\tilde{\alpha}(x,k(x))\,.
\ee
It follows that
\be
\nabla \tilde{\tilde{\alpha}}(x)= \left.\nabla_x \alpha(x,k)\right|_{k=k(x)} + \left. \partial_{k_i} \tilde{\alpha}(x,k)\right|_{k=k(x)} \, \nabla_x k_i (x) =  \left.\nabla_x \alpha(x,k)\right|_{k=k(x)}\,.
\label{from}
\ee
It is immediate that the function $\tilde{\tilde{\alpha}}$ will also satisfy $(\nabla \tilde{\tilde{\alpha}})^2=0$ and is a solution of the eikonal equation. Notice that as $\alpha(x,k)$ satisfies the same additive jump boundary condition (\ref{bc}) for all $k$, $\nabla_k \, \alpha(x,k)$ does not have jumps, and the same boundary condition will hold for  $\tilde{\tilde{\alpha}}(x)$.

In consequence, complete sets of solutions are generally not unique, and the particular set of eikonal solutions one chooses does not have physical significance.\footnote{This was misstated in \cite{arias2017anisotropic}.} However, the important point is that the entanglement temperatures are invariant under these  transformations. This follows directly from (\ref{from}). 
This equation implies that the gradient of the function at a point $x$ coincides with the gradient of a particular solution in the original set of solutions (the one with $k=k(x)$), at the same point $x$. The entanglement temperatures and their directions are precisely given by these gradients, eq. (\ref{14}). 
A simple example of the ET of Rindler space is presented in Appendix \ref{app1}. 

\subsection{Sources and metric}

Entanglement temperatures can be computed in an analogous manner for states created by sources in the Euclidean Lagrangian. Sources can be placed at $x^0\le 0$ to create a state different from the vacuum at $x^0=0$. The problem then contains the reflected sources at $x^0>0$ in the description of the density matrix by the path integral in the Euclidean plane. Then suppose we have a Lagrangian
\be
L= \frac{1}{2}(\partial \phi)^2+ J(x) {\cal O}(x)\,,
\ee
where the new term is symmetric under time reflections. If ${\cal O}$ has dimension $\Delta>d$ the path integral is ill-defined. For relevant operators, $\Delta < d$, the new terms appear in the equation of motion but will produce subleading corrections in the modular parameter $s$ for large $s$. Then, these terms will not affect the entanglement temperatures or the form of the high energy density matrix. For example, this is the case of the vacuum for massive fields, or more generally superrenormalizable theories, or coherent states. All these states share the same entanglement temperatures. For marginal operators $\Delta=d$ the applicability of the eikonal equations presumably depends on where the theory is asymptotically free. A clear case is when the perturbations source the stress tensor. In that case, it is natural to think of the source as the metric tensor. It is clear that the eikonal equations are turned to the generally covariant ones 
\be
g^{\mu\nu} (\partial_\mu \alpha) \, (\partial_\nu \,\alpha)=0\,. \label{58}
\ee
 This can indeed change the local temperatures. The metric can be non-trivial even at $x^0=0$ provided we have time reflection symmetry. The same holds for changes in the topology of the space. An example is a global thermal state where the topology is the one of a cylinder with periodic time direction. 

In the case of a non trivial metric,  $\partial_0 \alpha =A_0$ at $x^0=0$ gets combined as $\frac{2\pi}{A^2}\, A_0 P^0$ in the local density matrix. In terms of particle creation operators and momentum at a local inertial system eq. (\ref{esaa}) remains the same but we have to understand the inverse temperature as
\be
\beta(\vec{x},\hat{p})= g_{00}^{1/2}(\vec{x})\, \frac{2\pi}{A_0(\vec{x})}
=\frac{2 \pi}{|A(\vec x)|}=\frac{2 \pi}{|B(\vec x)|}\,.\label{ET}
\ee
It picks up a dependence on the metric analogous to Ehrenfest-Tolman effect $\beta\sim g_{00}^{1/2}$ for a thermal equilibrium state in a static metric.  The origin of this factor is the same, since $\beta$ settles the rate between proper time and modular time, as in the Ehrenfest-Tolman effect \cite{rovelli2011thermal}. 

\subsection{Conformal invariance}

Solutions to (\ref{uno}) and (\ref{dos}), or, equivalently (\ref{58}), are invariant under Weyl transformations of the metric $g\rightarrow \tilde{g}=\Omega^2 g$. So the same solutions of $\partial_\mu \alpha$ hold for both metrics. 
The entanglement temperatures (with respect to local inertial system excitations) change according to the scale factor as implied by (\ref{ET}), 
\be
\tilde{\beta}= \Omega \, \beta\,,\hspace{.7cm}    \tilde{T}= \Omega^{-1} \, T\,. 
\ee
This new temperature corresponds to the conformally transformed state for the new metric. The relative entropy for the transformed state and the transformed excitation is invariant because the energy of the excitation gets an opposite compensating factor in (\ref{eqq}). This is expected since the conformal transformation can be thought of as just a change of the coordinates in the description of a conformal theory, that cannot change the relations between algebras and states when properly identified.   

Therefore, the problem of understanding the high energy density matrix only depends on the conformal class determined by the metric, the region, and the UV conformal fix point of the theory.   

\subsection{Lorentzian eikonal equations and propagation of ET }
\label{Lorentz-ET}

The eikonal equations give the ET in vacuum at $x^0=0$ for a region $V$ based at $x^0$. However, we would like to understand the high energy terms in the modular Hamiltonian written at different Cauchy surfaces. This is also relevant to compute relative entropies with localized excitations at different spacetime points in the causal development of $V$.  

To understand the ET at $t\neq 0$  we need to extend the solutions of the wave equations determining the eigenvectors of the modular Hamiltonian from the Euclidean plane to the Minkowski one. This can be accomplished by gluing Euclidean space and Minkowski space at $x^0=0$.  
The equation for a massless scalar $\phi$ in the Minkowski region is again $\square \phi=0$ but with the Minkowski metric. In the eikonal approximation $\phi= e^{s \, \alpha_M(x)}$, $\alpha_M=a_M+i \, b_M$
  the eikonal equation is again $(\nabla \alpha_M)^2=0$, or equivalently (\ref{uno}) for the gradients $A_M=\nabla a_M$, $B_M=\nabla b_M$, but the scalar products are now Lorentzian.  

Let us call $\phi_E$ and $\phi_M$ to the Euclidean and Lorentzian solutions. At $x^0=0$ the matching conditions between solutions are 
\be
\vec{\nabla} \phi_M=\vec{\nabla} \phi_E\,,\qquad \partial_0 \,\phi_M= -i\, \partial_0 \,\phi_E\,. 
\ee 
These matching conditions give at $x^0=0$, taking into account (\ref{ot}), 
\be
A_M(x)=0\,,\qquad   B_M(x)=(A_E^0(x),B_E(x))\,, \qquad x^0=0\,.
\ee
This eliminates the real part in $\alpha_M$, and as a consequence, we have the Lorentzian eikonal equation
\be
B_M=\nabla b_M\,,\qquad B_M^2=0\,. 
\ee
This is, $b_m$ is a function whose gradient is everywhere a null vector. 
This null vector carries both the information of the direction in which the temperature is taken and its value given by 
\be\label{TempMod}
T= \frac{B_M^0}{2 \pi}=\frac{|\vec{B}_M|}{2\pi}\,.  
\ee

\begin{figure}[t]\centering
\includegraphics[width=.9\linewidth] {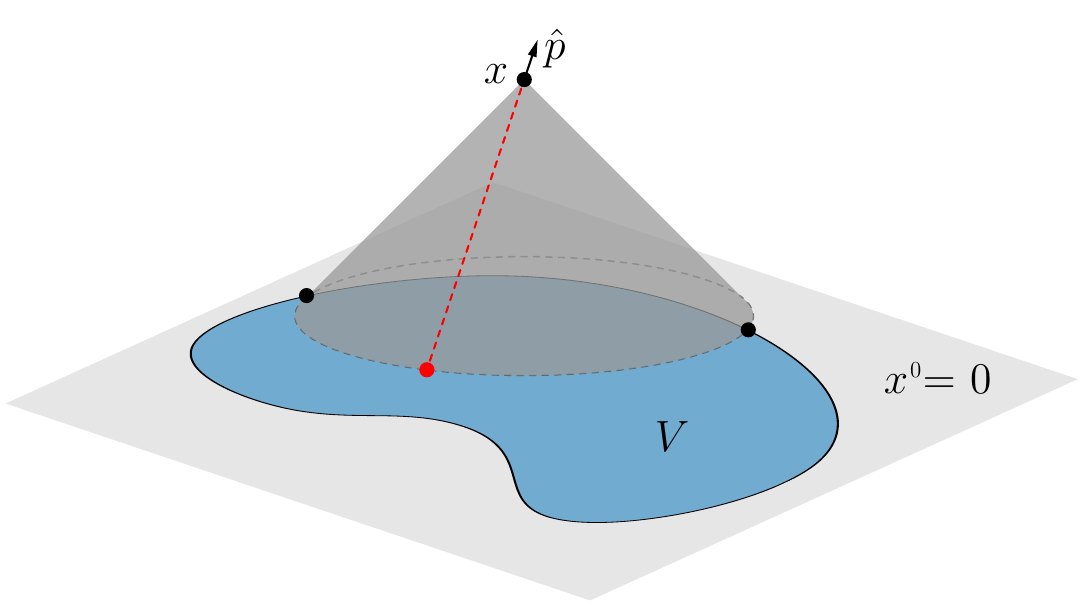}
\caption{The ET propagates ballistically from the $x^0=0$ surface where it is determined by the Euclidean eikonal equations. For example, the black dots correspond to directions in which the ET at $x$ diverges. This is the universal Rindler-like behavior of ET at the boundary of a system.}
\label{Fig:LBeta}
\end{figure}

The Lorentzian approach allows us to understand the covariance property of the ET  in an interesting light. We can actually define a temperature null vector as 
\be
T_\mu=(2\pi)^{-1}\, (B_M)_\mu\,.
\ee
From $T_\mu T^\mu=0$ we have
\be \label{L-EikoEqs}
\nabla_\mu (T_\nu T^\nu)= T^\nu \nabla_\mu T_\nu= T^\nu \nabla_\nu T_\mu=0\,,
\ee
where in the last step we used the fact, that $T_\mu$ is a gradient. Therefore, we conclude that the
temperature vector of a given solution to the eikonal equations is the tangent of affinely parametrized geodesics. In particular, it is parallel transported along the geodesic itself. As a result $T$ propagates ballistically in the direction determined by itself, i.e., it is the tangent of an affinely parametrized null geodesic. 
 In other words, ET are described by a choice of affine parameter for each null geodesic. 
This holds in general geometries and not only in flat space. This gives a complete set of values for the entanglement temperature vectors of any point in space-time in the causal development of $V$. Given a point $x$ and a direction $\hat p$, we just have to draw the corresponding geodesic back to the plane $x^0=0$ where we can read off the value given by the Euclidean solution, see Fig. \ref{Fig:LBeta}. In Minkowski space the parallel transport is trivial and we get the propagation law for the temperature vector, 
 \be
T(x+\lambda \,T(x))=T(x)\,. 
\ee

On the other hand, the Lorentzian propagation by itself does not give enough information to uniquely determine the ET in the Lorentzian context. The reason is essentially that the equation for the propagation does not determine the state, and the ET could in principle be fixed to arbitrary values on a Cauchy surface and then evolved ballistically. To get the vacuum ET we should select the vacuum state by continuing the manifold with an Euclidean section. The data at $x^0=0$ is then selected in this way. 
 
 Using the temperature vector, the relative entropy with a localized null excitation takes a remarkable expression. The momentum of the excitation and the temperature vector have to be taken in the same direction. The factor between the two null vectors turns out to be precisely the relative entropy, generalizing eq. (\ref{eqq}) :
\be
p^\mu =S_{\rm rel}\,\,\, T^\mu\,. \label{rela}
\ee
As $p^\mu$ is also an affinely parametrized tangent to the geodesic, $S_{\rm rel}$ is kept constant along the geodesic and has the interpretation of the ratio between affine parameters. 
The relative entropy is necessarily kept constant along the particle trajectory, because it is the relative entropy between two states, which cannot depend on the Cauchy surface where we choose to evaluate it \cite{Arias:2016nip}.

With the eikonal solutions extended to real time, it is possible to obtain the large modular energy part of the modular Hamiltonian written in an arbitrary Cauchy surface $\Sigma$ for the causal development of $V$. It also follows directly using (\ref{rela}) to calibrate the local terms:
\be
K_{\rm HE}=\int_{\Sigma} d\sigma\, \int d^{d-1}p\, \,\frac{p\cdot \hat n}{T(\hat{p})\cdot \hat n} \,\,   a_{\vec{p},\vec{x}}^\dagger \, a_{\vec{p},\vec{x}}\,, \label{esaab}
\ee
where $\hat n$ is the unit normal to the surface and the mode's spatial momentum and creation operators correspond to quantization in a local inertial system with spatial directions parallel to $\Sigma$.  

For special cases, the modular Hamiltonian is fully local (e.g., vacuum state for the Rindler wedge or a sphere in a CFT in Minkowski space), and has the expression 
\be
K=\int_\Sigma d\sigma \, \hat n^\mu \,T_{\mu\nu} \,\xi^\nu\,, \label{local}
\ee
where $\xi^\nu$ is a time-like future-directed conformal Killing vector. In this case, the relative entropy with a localized null excitation has the form
\be\label{Srel-killing}
S_{\rm rel}= - p_\mu \, \xi^\mu\,,
\ee
and $\xi^\mu$ the interpretation of an inverse temperature vector. From this expression, it follows that the temperature vectors are given in this case by
\be
T^\mu= \frac{p^\mu}{-p\cdot \xi}\,,\label{236}
\ee
where $p^\mu$ is any null vector determining the direction on the null cone. Plugging this expression in (\ref{esaab}) we can see the compatibility with (\ref{local}). Furthermore, one can check that this temperature vector satisfies the ballistic propagation (\ref{L-EikoEqs}) which follows from the conformal Killing equations:
\bea\label{conf-Killing}
\nabla_\mu \xi_\nu+\nabla_\nu \xi_\mu=\frac2d g_{\mu \nu} \nabla_\rho \xi^\rho\,.
\eea

It is worth noting that in the general case the high energy part of the modular Hamiltonian (\ref{esaab}) cannot be expressed by the integral of a field operator but its expression in terms of fields contains bilocal kernels \cite{Arias:2016nip}.

\subsection{Example: Entanglement temperature for Hawking radiation}
\label{app2}

\begin{figure}[t]\centering
\includegraphics[width=.65\linewidth] {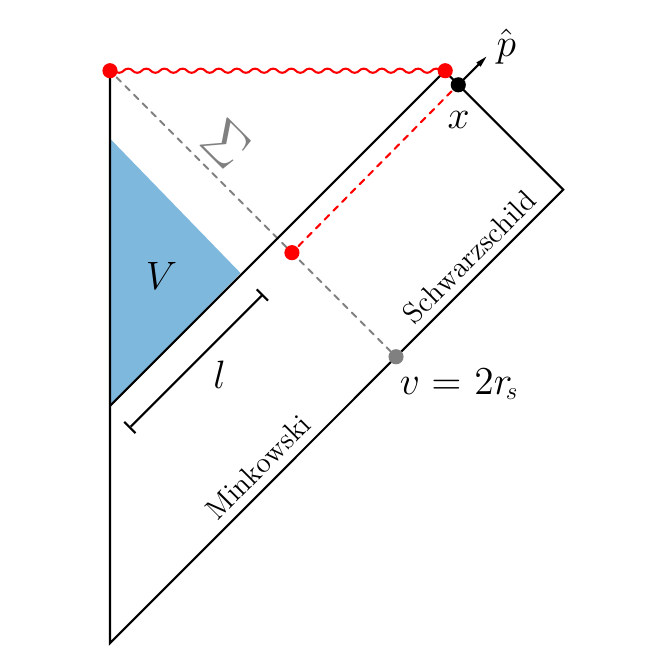}
\caption{Penrose diagram of the Vaidya metric. The entanglement temperature at a point $x$ in the asymptotic region is computed by first propagation back to the $v=2 r_s$ surface. There the ET is evaluated using the Killing vector corresponding to the modular Hamiltonian of vacuum in flat space corresponding to the light-blue region. The red dot marks $x_\Sigma$, where $\hat p_\Sigma$ is found by shooting a past null geodesic from $x$ in direction $\hat p$.}
\label{Fig:Vaidya}
\end{figure}

As an example of ET propagation, we compute ET for black hole formation and evaporation geometry described by a Vaidya metric
\be
ds^2=-\left(1-\frac{r_s }{r}\,\theta(v-2 r_s)\right)\, dv^2 +2 \,dv\,dr + r^2\,d\Omega^2\,.
\ee
The collapsing shell of radiation forms a null surface $\Sigma$ located at $v=2 r_s$. The Penrose diagram is shown in figure \ref{Fig:Vaidya}. We take the region as the complement of the double cone $r+|t-r_s|<r_s$, 
 with $t=v-r$, in the Minkowski region. The origin of the Minkowski coordinates is the tip of the black hole. This double cone is the part of the black hole that is inside the Minkowski region. The ET on the Minkowski region is given by the formula (\ref{236}), where $\xi$ is a conformal Killing vector keeping the double cone in itself. 
On the surface $\Sigma$ this conformal Killing vector is 
\be
\xi= \pi\, u_M \,(u_M- 2 r_s)/r_s\,\, \partial_{u_M}= -2\pi\,r\, (r-r_s)/r_s \,\,\partial_r \,,
\ee
where $u_M=t-r$ is the null Minkowski coordinate in the Minkowski region. Then, for any point $x$ on the exterior of the black hole and any null direction $p$ at this point, the ET vector is given by 
\be
T^\mu= \frac{p^\mu} {-p_\Sigma\cdot \xi}\,,
\ee
where $p_\Sigma$ is the parallel transport along the null geodesic determined by $x$ and $p$ back to the surface $\Sigma$. On the other hand, $p\cdot \xi_t$, where $\xi_t=\partial_v$ is the time translation Killing field of the Schwarzchild geometry, is a constant along geodesics. Then we can write
\be
T^\mu= -\left(\frac{p_\Sigma\cdot \xi_t}{p_\Sigma\cdot \xi}\right)_\Sigma\,\left(\frac{p^\mu} {p\cdot \xi_t}\right)_x\,.
\ee
For radial geodesics and points $x$ far away from the black hole this is 
\be
T= \frac{r_s}{4\, \pi \,r_\Sigma^2}\,\hat{p}\,, 
  \label{dd}
\ee
with $\hat{p}^\mu\equiv(1,1,0,0)$ in the  coordinates $\{t,r,\theta,\varphi\}$, and $r_\Sigma$ is given by the formula
\be
u_{BH}=2r_s- 2\left(r_\Sigma+r_s \log\left(\frac{r_\Sigma}{ r_s}-1\right) \right)\,,
\ee
where $u_{BH}=t-r$ in the coordinates of $x$, for $r\gg r_s$. 
Very fast after the formation of the black hole, $u_{BH}\gg r_s$, we have  
 $r_\Sigma\rightarrow r_s$, and (\ref{dd}) converges to a constant value given by the Hawking temperature
\be
T=\frac{1}{4\, \pi \,r_s}\,\hat{p}\,.\label{sss}
\ee

For $x$ in this asymptotic regime $t,r\rightarrow \infty$, $u\gg r_s$, only geodesics starting at $\Sigma$ for $r-r_s\ll r_s$ can reach $x$. 
 Then define the impact parameter 
\be
b=\left|\frac{\xi_\varphi\cdot p}{\xi_t \cdot p}\right|\,,
\ee
where $\xi_\varphi=\partial_\varphi$ is the Killing field of rotations.
This is conserved on geodesics. Starting from the shell, as is well known, only geodesics with $b<\frac{3 \sqrt{3}}{2} r_s$ will reach infinity. 
Geodesics for other values of $b$ will fall into the black hole. These geodesics start at $\Sigma$ being nearly radial geodesics, so the ET is the same one as for the radial case. In consequence, in the asymptotic limit, the only non-zero temperatures are for geodesics pointing in the black hole direction with an impact parameter less than $\frac{3 \sqrt{3}}{2} r_s$, and all have the same ET given by Hawking temperature (\ref{sss}).

We can also explore what happens if the region is the complement of a smaller double cone inside the Minkowski region of the black hole. As the region increases the temperature has to decrease. If we move the edge of the double cone away from the black hole horizon the ET in the asymptotic region will vanish because the inverse ET  $k_\Sigma\cdot \xi$  produced by the double cone on $\Sigma$ has a non-zero limit for $r\rightarrow r_s$, while the redshift factor $k_\Sigma\cdot \xi_t$ vanishes for late times. On the other hand, if we choose a double cone with the same past tip coinciding with the tip of the black hole, and a size $\Delta u_{BH}=l\in (0, 2 r_s)$, the inverse temperature conformal Killing vector produced by the double cone on $\Sigma$ near the horizon is
\be
\xi= - 2 \pi \, (r-r_s)\, \partial_r + 2\pi\, \frac{2r_s}{l} (2 r_s-l)\, \partial_v\,.
\ee
The result for asymptotic observers is again a uniform temperature on the disk of $b< \frac{3 \sqrt{3}}{2} r_s$, but with a smaller temperature  
\be
T=\frac{1}{4\, \pi \,r_s}\,\frac{l}{2 r_s}\,\hat{p}\,.\label{sss2}
\ee

\section{Rényi-0 and entanglement temperatures}\label{Sec:Renyi-O}

The vacuum density matrix, by definition, gives vacuum expectation values for the operators. Therefore it is not possible to test the high-energy thermal-like behaviour by simply studying operator expectation values. On the other hand, as we have recalled in the last section, the relative entropy with a localized excitation goes directly to examine this aspect of the reduced density matrix.

Another way in which the reduced density matrix will display its high energy sector is by considering powers or, equivalently, by considering states at different dimensionless 
temperatures with respect to the modular Hamiltonian,
\be
\rho_n= \frac{\rho^{n}}{\textrm{tr}\, \rho^n}=\frac{e^{-n K}}{\textrm{tr} e^{-n K}}\,.
\ee
This state now contains real excitations and a nonzero energy density. It gives a state that is singular at the boundary of the region. However, the expectation values of operators inside the region are well-defined, nontrivial, and can be studied.  
In the limit of high modular temperature, $n\rightarrow 0$, this density matrix will indeed reveal a highly excited gas of particles which is determined locally by the approximate high energy modular Hamiltonian (\ref{esaa}), that is completely determined by the ET. In the next section we will find this ``fluid'' obeys a relativistic Boltzmann equation with conserved current and stress tensor. 

Information quantities can be computed in this limit from the knowledge of the ET. 
In particular, the Rényi entropy is 
\be
S_n=\frac{1}{1-n}\,\log \tr \rho^n=\frac{1}{1-n} \big(\log Z(n)- n\log Z(1)\big)\,,
\ee
with the free energy
\be
\log Z(n)=\tr \,e^{- n K}\,.
\ee 
The limit of $n\rightarrow 0$ the Rényi entropy coincides with the free energy
\be\label{Rényi-zero}
S_n\sim \log Z(n)  \,, \hspace{1cm} n\ll 1\,.
\ee
We will be expressing the results in terms of the $ n\to 0$ limit of the Rényi entropy of the original vacuum density matrix.\footnote{For finite systems this limit of the Rényi entropy is called the ``Hartley entropy'' or ``max-entropy'', and it measures the rank of the density matrix. This is not applicable to the case of a QFT since this rank is always divergent, and in fact, we will find the limit of the Rényi entropy diverges for $n\rightarrow 0$. Hence, we prefer to use the term Rényi-$0$ instead of Hartley entropy.} It is also worth remarking that the geometric dependence of this quantity has the same general features as any other Rényi entropy, in particular, the entanglement entropy. That is, we have divergent area terms and subleading terms, and universal parts that can be isolated in different ways by understanding the geometric dependence of the divergent pieces. In particular, mutual Rényi entropies are always universal.

For an ordinary thermal state in a flat space and a CFT, the free energy has the form
\be
\beta\,F\equiv\log Z(\beta)= \sigma\,\frac{V}{\beta^{d-1}}\,, 
\ee
where $\sigma$ is a dimensionless coefficient that depends on the CFT, and $V$ is the volume. It is simply related to the thermal energy density (Stefan-Boltzmann law) or the thermal entropy density. For a free boson and fermion fields (per field degree of freedom) it is
\begin{align}
\sigma_B &= \frac{{\rm vol}(\mathbb{S}^{d-2})}{(2 \pi)^{d-1}}\, \int_0^\infty d\chi\, \chi^{d-2}\,(-)\, \log(1-e^{-\chi})=\frac{ \zeta (d) \Gamma
   \left(d/2\right)}{\pi ^{d/2}}\,,\label{sigmaB}\\\nn\\
\sigma_F &= \frac{{\rm vol}(\mathbb{S}^{d-2})}{(2 \pi)^{d-1}}\, \int_0^\infty d\chi\, \chi^{d-2}\,\log(1+e^{-\chi})=\frac{ \zeta (d) \Gamma
   \left(d/2\right)}{\pi ^{d/2}}\left(1-2^{1-d}\right)\,. \label{sigmaF}
\end{align}

The conjecture we would like to put forward is the following: The Rényi entropy for any QFT and a given spatial region $A$ in the $n\rightarrow 0$ limit, is proportional to $n^{-(d-1)}$ and the  Stefan-Boltzmann constant $\sigma$, as would be the case of the thermal entropy for a temperature $\sim n^{-1}$. Furthermore, it is proportional to a function $g(A)$ of the geometry of the region that characterizes the UV fix point and is universal for spheres.  In summary, the conjecture states 
\be\label{conjecture-g(A)}
S_n(A) \sim \frac{\sigma}{n^{d-1}}\,g(A)\,, \hspace{.7cm} \textrm{for }\, n\rightarrow 0 \,.
\ee 

For free UV fix points, we can provide a physical derivation of the above conjecture which allows us to find an explicit form for $g(A)$, as follows. First, we are interested in computing the following partition function
\bea
 Z_A(n)=\tr \rho_A^n=\tr e^{n K_A}
\eea
in the $n\to 0$  limit. This is a high-temperature limit, and thus this partition function can be approximated by the semi-classical phase space integral with the associated occupation number formula for a free bosonic or fermionic theory (depending on the corresponding free theory in the UV) and taken as its energy its local modular energy. Such local modular energies were studied in  \cite{arias2017anisotropic}, where the authors studied the local structure of the modular Hamiltonian in free QFTs and showed that for states with a large and localized energy excitation $|x, \vec{p}\,\rangle$, the expectation value of the ground state modular Hamiltonian scale with the energy of the excitation in a position and direction-dependent way.   This is, at high energies the modular Hamiltonian around a point $x$ is approximately given by, c.f. (\ref{esaab})
\bea\label{KA-local}
K_A(x)\approx \int d^{d-1}p\, \beta(x, \hat{p}) |\vec{p}\,| a_p^\dagger(x) a_p(x) \,.
\eea
In writing the above expression we used the ultra-relativistic approximation $E\approx |\vec{p}|$ since the energies involved in this regime are assumed to be much larger than the particle masses. The partition function $Z_A(n)$ in the $n\to 0$ limit is thus given
\bea\label{freedistrib}
\log Z_A(n)=\int_A d^{d-1}x \int \frac{d^{d-1} p}{(2\pi )^{d-1}}\(\pm\) \log\(1\pm e^{-n \beta(x, \hat{p})|\vec{p}\,|}\)\,,
\eea
where the sign $(\pm)$ depends on whether the free constituents are either fermions $(+)$ or bosons $(-)$. By appropriately re-scaling the momentum integral, namely making $\chi \to n \beta(x, \hat{p}) |\vec{p}|$ the above expression can be re-written as 
\bea
\log  Z_A(n)= \frac{1}{n^{d-1}}\int_A d^{d-1}x \int \frac{d^{d-2} \Omega}{\beta^{d-1}(x, \hat{\Omega})} \int_0^{\infty} \frac{d \chi\, \chi^{d-2}}{(2\pi )^{d-1}} \(\pm\) \log\(1\pm e^{-\chi}\)\,.
\eea
After one identifies the associated Stefan-Boltzmann coefficient, we arrive at the form  (\ref{conjecture-g(A)})
\begin{equation}\label{Rényi-zero-formula}
S_n(A)\approx \frac{\sigma}{n^{d-1}}g(A)\,  \quad {\rm with}\quad g(A)=\frac{1}{{\rm vol}{(\mathbb{S}^{d-2})}}\int_A d^{d-1}x \int \frac{d^{d-2} \Omega}{\beta^{d-1}(x, \hat{\Omega})}\,,
\end{equation}
where $g(A)$ is a coefficient that depends on the geometry $A$ and the ET. These in turn are obtained by solving a purely geometric problem.  We will compute $g(A)$ for various different geometries in the next subsections. 

A property that follows from (\ref{Rényi-zero-formula}) is the positivity of the Rényi-0 mutual information, 
\be
I_n(A,B)=S_n(A)+S_n(B)-S_n(A\cup B) \ge 0\,,\hspace{.7cm} n\ll 1\,. 
\ee
This is a known general property of this limit in finite systems \cite{Dam:2002}. Here it nicely follows from the monotonicity of relative entropy that implies $\beta(x,\hat{p})$ is a monotonically increasing function under increasing regions.  

In what follows, we will test the formula (\ref{Rényi-zero-formula}) in a series of different geometries where both,
 the local temperatures and the limit $n\to 0$ of the Rényi entropies associated with those geometries are known. In this way, via the explicit formulas for $g(A)$ in (\ref{Rényi-zero-formula}) we provide a series of non-trivial checks of this formula. 

\subsection{Spheres}

For a spherical entangling region of radius $R$, the entanglement temperatures are well known and given by 
\be
\beta(r)= 2 \pi \frac{(R^2-r^2)}{2 R}\,. 
\ee
As the above formula indicates, the local temperatures inherit the spherical symmetry from the geometry and are independent of the direction of the local excitation. The evaluation of the geometric factor $g(A)$ for the calculation of the Rényi-0 is straightforward:
 \bea
 g(A)=\int_A \frac{d^{d-1}x}{\beta^{d-1}(x)}
 =\frac{ {\rm vol}(\mathbb{S}^{d-2})}{\pi^{d-1}}\int_0^{1-\epsilon/R}\frac{ \xi^{d-2}d\xi}{\(1-\xi^2\)^{d-1}}\,,
 \eea
where we have re-scaled the radial coordinate as $\xi\to r/R$ and put a cut-off to control the area divergences. As usual, the explicitly divergent terms give rise to non-universal contributions to the Rényi-0 and then, we are interested in the universal logarithmic or finite piece contributions which are universal. The above integral can be expressed in terms of a Gauss hypergeometric function, but in order to extract such universal information we find it convenient to massage it further by doing the following change of coordinates
\bea
{\rm arctanh}\, u = \frac{2\xi}{1+\xi^2}\,,\quad  {\rm with }\quad u_{\rm max}={\rm arctanh}\(\frac{2\(1-\frac{\epsilon}{R}\)}{1+\(1-\frac{\epsilon}{R}\)^2}\)\approx -\log\(\frac{\epsilon}{R}\)\,.
\eea
In these coordinates, the coefficient $g(A)$ becomes
\bea
g(A)=\frac{ {\rm vol}(\mathbb{S}^{d-2})}{(2\pi)^{d-1}}\int_0^{u_{\rm max}} \sinh^{d-2}u\, du=\frac{\rm vol_{\rm reg}\(\mathbb{H}^{d-1}\)}{(2\pi)^{d-1}}\,,
\eea
where in the last line we have identified the integral with the volume integral of hyperbolic space. Thus, we can use well-known results for the universal terms in the regularized volume of hyperbolic space to extract the universal part of the geometric coefficient. For instance collecting results from \cite{Nakaguchi:2016zqi} we have 
 \bea
{\rm vol_{\rm reg}\(\mathbb{H}^{d-1}\)}=\frac{\pi^{d/2}}{\Gamma\(d/2\)}\times \left\{
\begin{array}{ll}
(-)^{\frac{d}2-1}\frac{2}{\pi}\log\(2R/\epsilon\) &  \quad {\rm for \,\, even \,\,}d\,,  \\
(-)^{\frac{d-1}2} & \quad {\rm for \,\, odd \,\,}d\,.
\end{array}
\right.
\eea
and from this, we get a final formula for the universal contribution to the Rényi-0 for spherical regions
\bea
\lim_{n\to 0}S_n=\frac{\sigma}{n^{d-1}}\frac{{\rm vol_{\rm reg}\(\mathbb{H}^{d-1}\)}}{(2\pi )^{d-1}}\,.\label{sphe}
\eea
For free theories, we can evaluate the Rényi-0 using the Stefan-Boltzmann coefficients (\ref{sigmaB}) and (\ref{sigmaF}). The results obtained in this way are in perfect agreement with the results of \cite{casini2010entanglement,Klebanov:2011uf}, for free bosons and fermions. One can read them off explicitly from equations (5.37)-(5.39) of \cite{Nakaguchi:2016zqi}, using the fact that the Rényi capacitance $C_n$ is related to the Rényi entropy via $S_n=d(d-1)C_n$, when $n\to 0$. 

In fact, (\ref{conjecture-g(A)}) holds for general CFT for spheres, giving (\ref{sphe}). The coefficient $g(A)$ is universal across CFTs for spheres because of the relation of the sphere density matrix with the thermal states in hyperbolic space \cite{Casini:2011kv}. At high temperatures, the curvature in hyperbolic space gives a subleading contribution to the thermal entropy, and this later is proportional to the flat space constant $\sigma$.   We can check this result in holographic theories. In this case, we need the value of the Stefan-Boltzmann coefficient in a holographic theory. For a $d-$dimensional CFT dual to Einstein gravity in $d+1$ we can extract this information, following standard thermodynamic relations in the holographic dual of a finite temperature CFT state. The procedure is standard,\footnote{Let us consider a holographic CFT at a finite temperature which is dual to a BH in an asymptotically AdS spacetime. The free energy of such a thermal state at high temperatures obeys the Stefan-Boltzmann law 
\bea
F=-\sigma_{holo} V T^d \quad {\rm or}\quad \log Z(T)=\sigma_{holo} V T^{d-1}\,.
\eea
This implies that the thermal entropy as a function of the temperature will have the form 
\bea\label{SB-coeff-holo}
S=-\(\frac{\partial F}{\partial T}\)=\sigma_{holo}\, d\,V T^{d-1}\,.
\eea
This quantity is easy to calculate in a holographic CFT as it is given by the entropy of its dual AdS Black hole 
\bea\label{HB-S}
S=\frac{L^{d-1}}{4G_N}\(\frac{4\pi}{d}\)^{d-1}VT^{d-1}\,.
\eea
Comparing the above result with (\ref{SB-coeff-holo}) allows us to deduce the value of the Stefan-Boltzmann coefficient in theories of Einstein gravity (\ref{sigma-holo}). 
} and leads to the relation 
\bea\label{sigma-holo}
\sigma_{holo} =\frac{L^{d-1}}{4G_N\, d}\(\frac{4\pi}{d}\)^{d-1}\,.
\eea
With this result in hand, we can write down the Rényi-$0$ in a holographic theory
\bea
\lim_{n\to 0}S_n=\frac{1}{n^{d-1}}\(\frac{2}d\)^{d-1}\frac{L^{d-1}}{4G_N\, d}{\rm vol}_{\rm reg}\(\mathbb{H}^{d-1}\)\,.
\eea
As expected, this result is in perfect agreement with the Rényi computations of \cite{Hung:2011nu} in the aforementioned limit.

From the universality of the coefficient $g(A)$ for spheres it follows the universality of $g(A)$ for regions with arbitrary boundaries on the null cone. For this type of regions, the Renyi entropies depend geometrically in a universal way on the boundary, and the coefficient is calibrated by the one of spheres \cite{casini2018all}.  

\subsection{Multi-intervals}\label{Sec3Multi-interval}

Let us consider the ground state of a $2d$ QFT reduced onto $N$ disjoint intervals, say $\cup_{i=1}^N I_i$ with $I_i=[l_i, r_i]$. The local temperatures associated with this geometry are known and were studied in \cite{Arias:2016nip,arias2017anisotropic}. The eikonal equations for the vector fields 
$A=\nabla a$ and $B=\nabla b$
characterizing the entanglement temperatures, see \eqref{ABdef}, adopt the form
\bea
\(\partial_x a\)^2+\(\partial_y a\)^2=\(\partial_x b\)^2+\(\partial_y b\)^2,\quad \partial_x a\, \partial_x b+\partial_y a\,\partial_y b=0\,.
\eea 
where $\alpha=a+ib$.  These equations are equivalent to the Cauchy-Riemann equations:
\begin{align}
\partial _x a&=\pm \partial_y b \,,\nonumber\\
\partial _y a&=\mp \partial_x b\,,
\end{align}
whose solutions are arbitrary holomorphic and anti-holomorphic functions on the complex plane.  The boundary conditions (\ref{bc}) on $a$ and $b$ across the cuts, together with regularity at infinity fix the solutions to 
\bea
\alpha(z)=i \log\(\prod_{i=1}^N\frac{z-l_i}{r_i-z}\)=a(x,y)+i\, b(x,y)\,,
\eea 
and $\alpha(\bar{z})$ its complex conjugate. The local temperatures are computed on the $t_E=y=0$ slice which is identified with the associated local temperatures in Minkowski signature on $t_M=0$ and are given by 
\bea
\beta(x)=\frac{2\pi}{|\nabla a(x,0)|}=2\pi\(\sum_{i=1}^N \[\frac{1}{x-l_i}+\frac{1}{r_i-x}\] \)^{-1}\,.
\eea
We can compute the Rényi-$0$ from the local temperatures of an arbitrary number of intervals, namely $A=\cup_i I_i$ and $I_i=[l_i, r_i]$. For a region at $t=0$ in $d=2$ the local temperatures do not depend on the direction of the probing localized perturbation, and thus the formulas for the geometric factor $g(A)$, (\ref{Rényi-zero-formula}) reduces to\footnote{One can see that \eqref{Rényi-zero-formula} reduces to (\ref{d=2-LT}) by putting $d=2$. This is, the angular integral reduces to two terms corresponding to the left and right directions, and the volume of a 0-sphere equals 2, i.e. ${\rm vol}{(\mathbb{S}^{0})} \equiv 2$.\label{footd2}} 
\bea\label{d=2-LT}
g(A)=\int_A \frac{dx}{\beta^{d-1}(x)}=\frac{1}{2\pi}\sum_{i=1}^N \int_A\(\frac{dx}{x-l_i}+\frac{dx}{r_i-x}\)\,.
\eea
For each $i-$th term, we separate the integral over $A$ into the piece that contains that region and the rest. The integral on $I_i$ needs a regulator, thus
\begin{align}
g(A)&=\frac{1}{2\pi}\sum_{i=1}^N \int_{l_i+\epsilon}^{r_i-\epsilon}\Big(\frac{dx}{x-l_i}+\frac{dx}{r_i-x}\Big)\nonumber\\
&=\frac{1}{\pi}\(\sum_{ij}\log|r_j-l_i|-\sum_{i<j}\log|r_j-r_i|-\sum_{i<j}\log|l_j-l_i|-N \log \epsilon\)\,.
\end{align}
The Rényi-$0$ is  
\bea\label{Rényi0-MIntervals}
S_n(A)\approx \frac{\sigma}{n\, \pi}\(\sum_{ij}\log|r_j-l_i|-\sum_{i<j}\log|r_j-r_i|-\sum_{i<j}\log|l_j-l_i|-N \log \epsilon\)\,.
\eea
For free fermions in $d=2$, $\sigma_f=\pi/6$ and the above formula reduces to the $n\to 0$ limit of the Rényi entropy formula for free fermions, see for example equation (177) of \cite{Casini:2009sr}. 
It is plausible that this formula should apply to any $2$d CFT because the ET are universal. A similar general check can not be carried out for the free scalar since their Rényi entropies are not known for an arbitrary number of intervals. However, they are known for two intervals. 

Let us then consider a chiral scalar field and the case of two disjoint intervals of lengths $l_A$ and $l_B$ separated by a distance $D$.  From (\ref{Rényi0-MIntervals}) we can compute the associated Rényi-$0$ mutual information between the intervals, which is given by 
\bea\label{2IntMut}
I^{s}_n(A,B)\sim -\frac{1}{12\,n} \log\( 1-\eta\)\,,\qquad\qquad \eta=\frac{l_A \, l_B}{(D+l_A) (D+l_B)}\,,
\eea
where $\eta$ is the cross ratio obtained from the endpoints of the intervals.
Notice that we have used $\sigma_f=\sigma_s=\pi/12$ and multiplied by an additional factor $1/2$ since we are considering only one chirality. 
The Rényi mutual information of two intervals in this theory is given by \cite{Arias:2018tmw}:
\bea\label{mutual-chiral-scalar}
I^s_n(\eta)=-\frac{n+1}{12\,n} \log\(1-\eta\)+U_n(\eta)
\eea
where the function $U_n(\eta)$ is 
\bea\label{LongDistExample}
U_n(\eta)=\frac{i}{2(n-1)} \int_0^{+\infty}ds \[ \coth(\pi s)-\coth(\pi s/n)\] \log\(\frac{\,_2F_1(1+i \frac{s}{n},-i\frac{s}{n};1;\eta)}{\,_2F_1(1-i \frac{s}{n},i\frac{s}{n};1;\eta)}\)\,. 
\eea
 This expression is hard to analyze for arbitrary values of $n$. In particular, the $n\to1$ limit was considered in \cite{Arias:2018tmw} where the expected long-distance behavior for the mutual information was recovered. Here, we are interested in studying the $n\to 0$ limit of $U_n(\eta)$.  First, let us notice that the integrand is well-defined everywhere on the half-real line. In particular, the apparent pole at $s=0$ in the hyperbolic cotangents is canceled by a simple zero in the log function. The integrand is everywhere bounded and decays exponentially at $s\to \infty $.

To carry out the above integral, we separate it into two pieces as 
\bea
U_n(\eta)=U^{<s_0}_n(\eta)+U^{>s_0}_n(\eta)\;.
\eea
For concreteness, let us take $s_0\sim {\mathcal O}(1)$ and estimate first the contribution of the lower part of the integral 
\bea
U^{<s_0}_n(\eta)=\frac{i\, n}{2(n-1)} \int_0^{n s_0}ds \[ \coth(n \pi s)-\coth(\pi s)\] \log\(\frac{\,_2F_1(1+i s,-i s;1;\eta)}{\,_2F_1(1-i s,is;1;\eta)}\) \,,
\eea
where we have re-scaled the integral by the change of variable $s\to ns$. In this form, the new integration variable runs over a small region $0<s<n s_0$ in the $n\to 0$ regime. Since the above integrand is bounded and differentiable at $s=0$, we conclude that  $U^{<s_0}_n(\eta)\to 0$ as $n\to 0$. 

The upper part of the integral is now
\bea
U^{>s_0}_n(\eta)=\frac{i}{2(n-1)} \int_{s_0}^{+\infty}ds \[ \coth(\pi s)-\coth(\pi s/n)\] \log\(\frac{\,_2F_1(1+i \frac{s}{n},-i\frac{s}{n};1;\eta)}{\,_2F_1(1-i \frac{s}{n},i\frac{s}{n};1;\eta)}\) \,.
\eea
In this region the integration variable $s_0<s<\infty$, since $s_0\sim {\mathcal O}(1)$, the ratio $s/n \to \infty$ when $n\to 0$ for all $s$. In that regime, the argument of the above log function simplifies drastically. The relevant asymptotic behavior can be derived from the integral representation of the Gaussian hypergeometric functions, by using the stationary phase approximation. This was done  in \cite{Arias:2018tmw}, where the authors found 
\begin{equation}
    \lim_{s/n \to \infty}i \log \(\frac{\,_2F_1(1+i \frac{s}{n},-i\frac{s}{n};1;\eta)}{\,_2F_1(1-i \frac{s}{n},i\frac{s}{n};1;\eta)} \)= i \log\(1-2\eta+ i2\sqrt{\eta (1-\eta)}\)=-\, \arccos\(1-2\eta\)\,.\nn
\end{equation}
Likewise, we can replace $\coth(\pi s/n)\to 1$. These approximations are exact in the $n\to 0$ limit, but for finite $n$ they are still useful as they provide an upper bound for $U^{>s_0}_n(\eta)$ due to the monotonicity properties of the approximated functions. We thus have
\bea
U^{>s_0}_n(\eta)=\frac 12 {\, \arccos\(1-2\eta\)}\left(s_0 +\log\left(\frac{1}{2}{\rm csch}( \pi s_0) \right) \right)\,,
\eea
where we have safely taken the $n\to 0$ limit. The coefficient that depends on $s_0$ is finite and small for $s_0\sim 1$ as was our choice and therefore, the Rényi-0 mutual information is dominated by the first term in (\ref{mutual-chiral-scalar}). We conclude that in the $n\to 0$ limit we recover the correct result \eqref{2IntMut} as expected, i.e.
\bea
I^s_n(\eta)=-\frac{1}{12\,n} \log\(1-\eta\)\,.
\eea

\subsubsection{Long-distance limit}

In the long-distance limit, the mutual information simplifies and can be computed exactly. However, the long-distance limit of R\'enyi entropies and the $n\to 0$ limit do not commute. This has to be so because the long-distance limit is dominated by the model-dependent lowest dimensional operator, while the $n\to 0$ limit is determined by the universal ET.  

Let us consider an explicit instance of these non-commutative limits in the example above. Consider again \eqref{LongDistExample}. For fixed arbitrary $n$, we can take the long-distance $\eta \ll 1$ limit inside the integrand. 
In that case, we have
\bea
 \log\(\frac{\,_2F_1(1+i \frac{s}{n},-i\frac{s}{n};1;\eta)}{\,_2F_1(1-i \frac{s}{n},i\frac{s}{n};1;\eta)}\)
 \approx -2 i\frac{s}{n}\eta - i \(\frac{s}{n}-\frac{s^3}{n^3}\)\eta^2+\cdots
\eea
Notice, that this expansion comes with polynomials in the ratio $s/n$, and therefore after integration one should be careful that the resulting series expansion is well-defined for the values of $n$ we are considering. In fact, the series in $\eta$ can be integrated term by term leading to 
\bea
U_n(\eta)=-\frac{(n+1)}{12 n}\eta -\frac{(n+1)(9n^2-1)}{240 n^3}\eta^2+{\cal O}(\eta^3)+\cdots
\eea
In this expression, we can see explicitly the failure of the expansion in the $n\to 0$ limit, as each term will have a larger and larger divergence in $1/n$ for a small but fixed $\eta$.

\subsection{Strip}

\begin{figure}[t]\centering
\includegraphics[width=.75\linewidth] {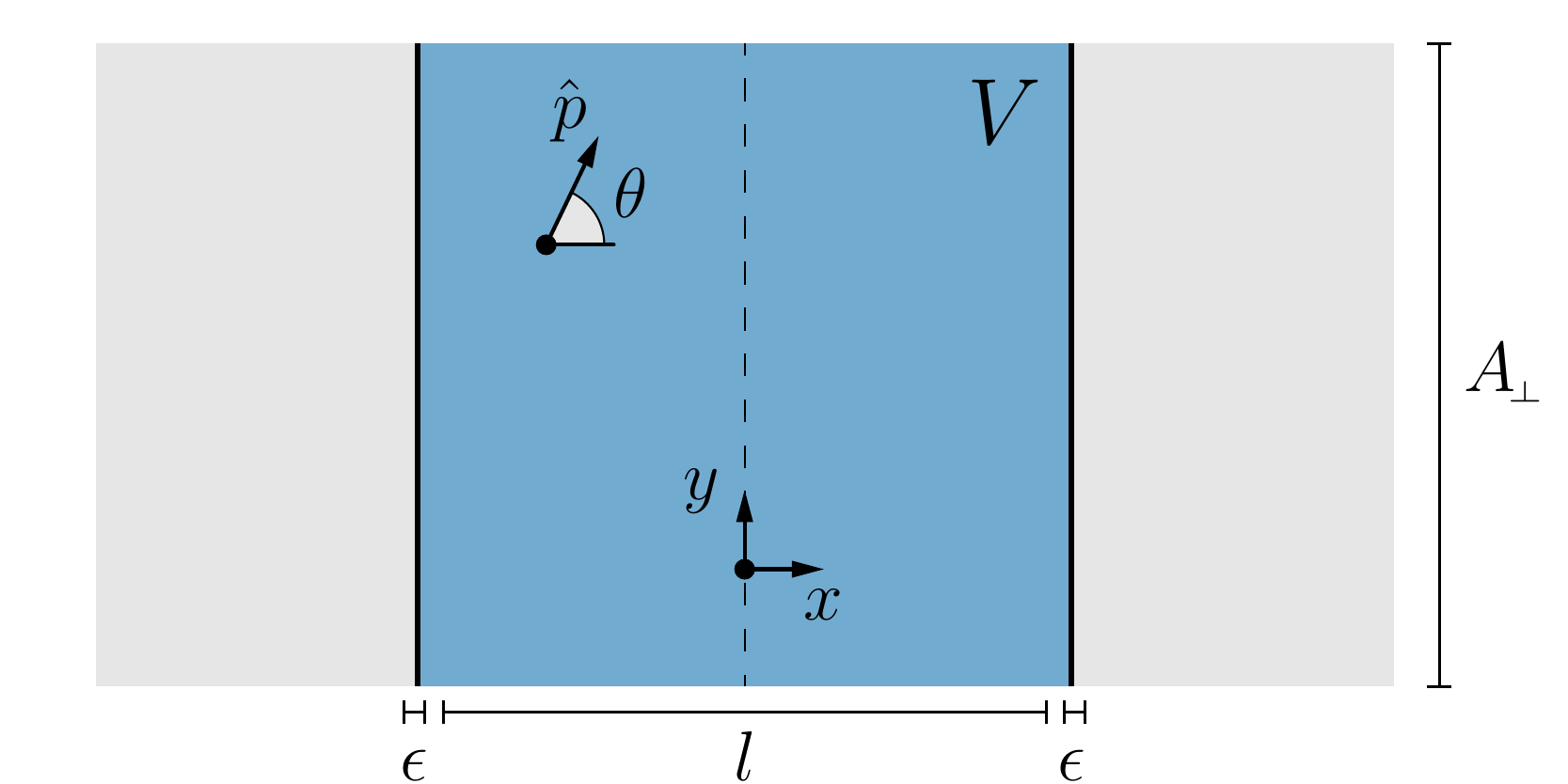}
\caption{ The strip for general $d$ consists of two parallel walls located at $x=\pm l/2$ that extend all over the perpendicular coordinates represented collectively by $y$ in the figure. The time coordinate points outwards from the figure. The $\theta$ angle is defined for any point as the angle between $\hat p$ and $x$. The $\epsilon$ regulators, at $x=\pm (l/2-\epsilon)$ are also shown. }
\label{Fig:Strip}
\end{figure}

The strip geometry is another example in which the Rényi entropies have been computed exactly for all non-integer values of the Rényi parameter for free fields. This geometry is characterized by its width $l$ and transverse area, which although infinite in size we regularized to a large but finite value $A_\perp$, see Fig. \ref{Fig:Strip}. The Rényi entropies for this geometry have the form
\bea
S_n\sim \alpha_n\,\frac{A_\perp}{\epsilon^{d-2}} - \kappa_n \,\frac{A_\perp}{l^{d-2}}\,, 
\eea
where the first term is cut-off dependent and thus non-universal, while the second term is universal. For free fields this was obtained by dimensional reduction in \cite{Casini:2005zv, Casini:2009sr}. In the next subsection, we will review this calculation as applied to the $n\to 0$ limit. 

\subsubsection{Direct computation of Rényi-$0$}

The calculation of $S_n$ presented in \cite{Casini:2005zv, Casini:2009sr} is done using the replica trick.
In this setup one is interested in computing the partition function of $n$ copies of the theory with some non-trivial monodromy around the boundaries of the entangling region. For free fields, one can choose a basis of free fields
that diagonalizes the monodromy into a simple multiplicative boundary condition $e^{i 2\pi k/n}$ where $k$ labels the different copies. Thus, the Rényi entropy becomes a sum over the different diagonalized copies. For scalars we have
\bea
S_n=\frac{1}{1-n}\sum_{k=0}^{n-1}\log Z(e^{i 2\pi \frac kn})\,,\label{36}
\eea
 where  $\log Z(e^{i 2 \pi  \frac{k}n})$ is the free energy associated with each diagonalizing free field (with the plane free energy subtracted). The partition function associated with a strip geometry for a free field can be obtained by dimensional reduction from the partition function of a massive field in one interval in $d=2$.
 In this way, one gets \cite{Casini:2009sr}
\begin{equation}
\kappa_n = h(d) \,\int_0^\infty dy\, y^{d-3} \, c_n(y) \,,\label{ka}\qquad\qquad
h(d)^{-1} = (4\pi)^{\frac{d-2}{2}} \Gamma( d/2)\,.
\end{equation}
Here $c_n(y)=R\, \frac{dS_n}{d R}$, where $S_n$ is the Rényi entropy of an interval of size $R$ in $d=2$ and a free massive field, and $y=m R$.  
In the same way as (\ref{36}) for the higher dimensional case, one has a sum over copies for the $d=2$ fields,
 \be
 c_n(y)= \frac{1}{1-n}\, \sum_{k=1}^{n-1}\,w_{\frac{k}{n}}(y)\,, \qquad\qquad \omega_a(y)=R \,\frac{d}{dR}\log Z(e^{i 2 \pi a})\,.
 \label{34} 
 \ee
The functions $c_n(y)$ can be obtained by integrating solutions of a Painlev\'e equation \cite{Casini:2005zv, Casini:2009sr}. 

 For the scalar, the problem chooses naturally $a\in (0,1)$, and $\omega_0(y)=\omega_1(y)=0$. There is a symmetry $a\rightarrow 1-a$ in $\omega_{a}(y)$. In order to continue (\ref{34}) analytically for all $n>0$ we use a residue formula
 \be
 c_n(y)=\frac{(2 \pi i)^{-1}}{1-n} \,\int dz\, \sum_{k=1}^{n-1} \frac{1}{z-e^{i 2 \pi k/n}} \,\omega_{\frac{\log z}{2\pi i}}(y)= \frac{(2 \pi i)^{-1}}{1-n}\,\int dz\,n \frac{z^{n-1}}{z^n-1}\omega_{\frac{\log z}{2\pi i}}(y)\,.\nn
 \ee
The contour encloses the unit circle where all poles for $n$ integer appear but has to enclose also the real negative axis, where there is a cut of the integrand, in order to be an analytic continuation for non-integer $n$, see Fig. 4 of \cite{Casini:2009sr}. For $n\le 1$ there are no more poles on the unit circle and only the cut around the real negative axis remains; $\omega_{\frac{\log z}{2\pi i}}(y)$ has no problems around this line. Changing the integration variable from $z\to b$, via $\frac{\log z}{2\pi i}=\frac12+ i b$, where the negative real line on the $z$ plane corresponds to real $b$, and taking into account the symmetry $b\to -b$, we get  
\begin{align}
c_n(y) &=\frac{1}{n-1}\,\int_0^\infty db\,\frac{n\, \sin(n\pi)}{\cos(n \pi)-\cosh(2 b n \pi)}\, \tilde{\omega}_{ b}(y)
\,, \hspace{1.2cm} n<1. \label{ll}     
\end{align}
We have defined $\tilde{\omega}_b(y)=-\omega_{1/2+ib}(y)$.
The function $\tilde{\omega}_{b}(y)$ is given by the equations
\begin{align}
\tilde{\omega}_{b}(y) &= \int^\infty_y  dt\, t\, u_b^2(t)\,, \label{39}\\
u_b''+\frac{1}{t}\, u_b' &=  \frac{u_b}{1+u_b^2} \, (u_b')^2 +u_b \, (1+u_b^2)-\frac{4 \,b^2}{t^2}\,\frac{u_b}{1+u_b^2}\,,\label{40}\\
u_b(t) &\rightarrow  \frac{2}{\pi}\, \cosh(\pi \,b)\, K_{i 2 b}(t)\,, \hspace{.7cm} t\rightarrow \infty\,.\label{41}
\end{align}
In particular, the entanglement entropy, $n\rightarrow 1$ is given by the known formula 
\be
c_1(y)=\int_0^\infty db\,\frac{\pi}{2 \cosh^2( b \pi)}\, \tilde{\omega}_{b}(y)\,,
\ee
and all Rényi entropies for $n<1$ depend upon the same functions $\tilde{\omega}_b(t)$.

The $n\rightarrow 0$ limit should be taken carefully. For arbitrary $n>0$, we can change the integration variable in (\ref{ll}) to $\tilde b=n b$, which gives  
\be\label{cny}
c_n(y)\sim \int_0^\infty d\tilde{b}\,\frac{n \pi}{2\,\sinh^2(\pi \,\tilde{b})}\, \tilde{\omega}_{ \tilde{b}/n}(y)\,.
\ee
In the $n\to 0$ limit, and for any finite $\tilde b$, we are looking at $\tilde\omega_b(y)$ built out of the asymptotic large $b=\tilde b/n$ solutions to (\ref{40}) in (\ref{39}). The differential equation (\ref{40}) has a closed simple solution at large $b$ which is naturally obtained after re-scaling $t\to b\, \tilde{t}$. This was studied in \cite{arias2017anisotropic}, where it was found that 
\begin{equation}\label{asymp-ubt}
\lim_{b \rightarrow \infty } u_{b} (b\,\tilde{t})=
\left\{
\begin{array}{ccc}
\sqrt{\frac{2}{{\tilde{t}}}-1} & \text{ for } & \tilde{t}>2 \, \\
0 & \text{ for } &  \tilde{t}>2\,. 
\end{array}
\right.
\end{equation}
Therefore, in the large $b$ regime, we have from (\ref{39}) and (\ref{asymp-ubt})
\bea
\tilde{\omega}_{b}(y)\sim  b^2\int^\infty_{y/b}  d\tilde{t}\, \tilde{t}\, u_b^2(b\, \tilde{t})=b^2\int_{y/b}^2  d\tilde{t}\, \tilde{t}\(\frac{2}{\tilde{t}}-1\)\Theta\[2-\frac yb\]=2\(b-\frac y2\)^2\Theta\[2-\frac yb\]\,,
\eea
where $\Theta[x]$ is the step function, which equals $1$ for a positive argument and $0$ otherwise.
With this result we can evaluate $\kappa^s_n$ using (\ref{cny}), and (\ref{ka})
\begin{align}
\kappa^s_n &\sim  h(d) \,\int_0^\infty dy\, y^{d-3} \,\int_0^\infty d\tilde{b}\,\frac{n \pi}{\sinh^2(\pi \,\tilde{b})}\, \(\frac{\tilde{b}}{n}-\frac y2\)^2\Theta\[2-\frac {y\, n}{\tilde b}\] \\
& \sim  \frac{h(d)}{n^{d-1}} \,\int_0^\infty d\tilde{b}\,  \frac{\pi \, \tilde{b}^{d}}{\sinh( \pi \,\tilde{b})^2}\int_0^\infty d\tilde y\, \tilde y^{d-3}\(1-\frac {\tilde y}2\)\Theta\[2-\tilde{y}\] \,,
\end{align}
where in the last line, we re-scaled the $y$ integral as $y\to \tilde{b}\, \tilde{y}/n$. This expression has the expected behavior $\sim n^{-(d-1)}$. Carrying out the above integrals, one gets 
\bea \label{kappa-strip-s}
\kappa_n^s\sim \frac{1}{n^{d-1}}\frac{\Gamma\(\frac{d-1}2\)\zeta(d)}{(d-2) \pi^{\frac{3d-1}2}}\,.
\eea
For the fermion a similar calculation holds. The formula (\ref{ll}) is now 
\be
c_n(y)
=\frac{1}{1-n}\,\int_0^\infty db\,\frac{n\, \sin(n\pi)}{\cos(n \pi)+\cosh(2 b n \pi)}\, \tilde{\omega}_{ b}(y)
\,, \hspace{.7cm} n<1. 
\ee 
The function $\tilde{\omega}_b(y)$ satisfies the same differential equations (\ref{39}), (\ref{40}), but now $u_b(t)$ has a different boundary condition
\be
u_b(t) \rightarrow  \frac{2}{\pi}\, \sinh(\pi \,b)\, K_{i 2 b}(t)\,, \hspace{.7cm} t\rightarrow \infty\,.
\ee
In the large $b,t$ limit, however, we have the same limit as the $u_b$ function for the scalar. The only difference with the scalar is that now the small $n$ limit gives 
\be\label{kappa-strip-f}
\kappa^f_n\sim  \frac{h(d)}{n^{d-1}}\, \frac{\pi\, 2^{d-1}}{d(d^2-3 d+2)}\,\int_0^\infty d\tilde{b}\,  \frac{\tilde{b}^{d}}{\cosh( \pi \,\tilde{b})^2}=\left(1-\frac{1}{2^{d-1}}\right)\kappa^s_n\,.
\ee
We see that, as expected, the ratio between the universal coefficients of the Rényi entropies in the $n\to 0$ limit for bosons and fermions is the same as the one between (\ref{sigmaB}) and (\ref{sigmaF}) for the Stefan-Boltzmann constants.

\subsubsection{Rényi-0 from local temperatures}

Let us take a strip defined to lie within the interval $|x|\leq 1/2$, and extend without bound along the extra dimensions say, $\{y^i\}$, with $i\in \{1,\cdots ,d-2\}$. The local temperatures are homogeneous and isotropic with respect to $\{y^i\}$ and thus, they can only depend on $x$ and the azimuthal angle $\theta$, with respect to the positive $x$ direction.  The local temperatures for the strip were studied in detail in \cite{arias2017anisotropic} and as shown there, they have a rich structure that interpolates between two limiting cases, the Rindler like, and the interval-like like which are separated by the critical curve
 \bea
\sin\theta=1-2|x|\,.
\eea
The ET for the strip are
\begin{equation}\label{beta-strip2}
\beta(x,\hat p)\equiv
\begin{cases}
\pi(1-2 |x|)& \text{ for }  \sin\theta > 1-2|x|\\
\frac\pi2 \frac{\cos^2\theta-4x^2}{1-|\sin\theta|} & \text{ for }  \sin\theta<1-2|x|
\end{cases}\;,
\end{equation}
i.e. the region $\sin\theta>1-2|x|$, has the same local temperatures than Rindler and $\sin\theta<1-2|x|$, has local temperatures which are described by a different function 
that reduces to the ones of the interval when $\theta \to 0$, and characterizes the non-trivial geometric dependence of the strip. Formula (\ref{beta-strip2}), differs slightly from the one presented in \cite{arias2017anisotropic}, but agrees exactly within $\sin\theta<1-2|x|$. 

Given the explicit formulas for $\beta$ we can now proceed to calculate the geometric coefficient $g(A)$ where $A$ is a strip. This requires computing 
\bea
g(A)=\frac{1}{{\rm vol}{(\mathbb{S}^{d-2})}}\int_A d^{d-1}x \int \frac{d^{d-2} \Omega}{\beta^{d-1}(x, \hat{\Omega})}\,.
\eea
The integral over $A$ separates between the coordinates parallel and perpendicular to the strip. The local temperatures depend only on the azimuthal angle which we denoted as $\theta$, therefore
\begin{align}
g(A)&=\frac{A_\perp}{{\rm vol}{(\mathbb{S}^{d-2})}}\int_{-\frac12}^{\frac12} dx \int_0^\pi  d\theta \sin^{d-3}\theta \int_{\mathbb{S}^{d-3}}\frac{d^{d-3} \Omega}{\beta^{d-1}(x, \theta)} \nonumber\\
&=\frac{{\rm vol}{(\mathbb{S}^{d-3})}\,A_\perp}{{\rm vol}{(\mathbb{S}^{d-2})}}\int_{-\frac12}^{\frac12} dx \int_0^\pi  \frac{\sin^{d-3}\theta  }{\beta^{d-1}(x, \theta)}\, d\theta \,.
\end{align}
We will start by evaluating the finite contribution that comes from the part of the strip whose local temperature equals the one of Rindler, which is delimited by $\sin\theta<1-2|x|$. In this region, the integrand is independent of $\theta$ and thus, it is convenient to carry out the integral over $x$ first 
\bea
g_R(A)=\frac{4\,{\rm vol}{(\mathbb{S}^{d-3})}\,A_\perp}{{\rm vol}{(\mathbb{S}^{d-2})}}\int_{\arcsin(2\epsilon)}^{\pi/2}   d\theta\, \sin^{d-3}\theta\, \int_{\frac{1-\sin\theta}{2}}^{\frac12-\epsilon}\frac{ dx}{\beta^{d-1}_R(x)}\,,
\eea
where the factor of $4$ comes from the separation of the integration region into four parts which are related by symmetry.
We put a regulator on the $x$ integral which also regulates the angular integral through the curve which delineates the Rindler region. The integral over $x$ gives 
\bea
 \int_{\frac{1-\sin\theta}{2}}^{\frac12-\epsilon}\frac{ dx}{\beta^{d-1}_R(x)}=\frac{1}{2(d-2)\pi^{d-1}}\(\frac{1}{\(2\epsilon\)^{d-2}}-\frac{1}{\sin^{d-2}\theta}\)\,,
\eea
and for the full integral we obtain
\begin{align}\label{g-rindler-result}
g_R(A)&=\frac{2}{(d-2)(2\pi)^{d-1}}\frac{A_\perp}{\epsilon^{d-2}}+\frac{\Gamma\(\frac{d-1}2\)}{\Gamma\(\frac d2\) \pi^{d-\frac{1}{2}}}\frac{A_\perp}{l^{d-2}}\log\(\frac{2\epsilon}{l+\sqrt{l^2-4\epsilon^2}}\) \nonumber\\
&\qquad\qquad\qquad -\frac{\Gamma\(\frac{d-1}2\)}{(d-2)\Gamma\(\frac d2\) \pi^{d-\frac{1}{2}}}\frac{A_\perp}{l^{d-2}}\,_2F_1\[\frac12,\frac{d-2}2,\frac d2;4\(\frac{\epsilon}{l}\)^2\]\,,
\end{align}
where we have recovered the length scale of the strip $l$. 
Let us analyze the divergent structure of this contribution. The first term gives the expected non-universal area law divergent piece. The second term gives a  $\log(\epsilon/l)$ divergent finite piece contribution in the $\epsilon/l \to 0$ limit. We know this universal contribution should be canceled by the contribution of the other solution. The final term, the term proportional to the Hypergeometric function, gives a constant universal contribution in the $\epsilon/l \to 0$ limit. 

Next, we evaluate the contribution that comes from the part of the strip whose local temperatures are given by (\ref{beta-strip2}). Integrating first over the angles the resulting integral is
\bea\label{g-strip}
g_S(A)=\frac{4\,{\rm vol}{(\mathbb{S}^{d-3})}\,A_\perp}{{\rm vol}{(\mathbb{S}^{d-2})}}\int_{0}^{\frac12-\epsilon} dx\int_{0}^{\arcsin(1-2x)}   d\theta\, \frac{ \sin^{d-3}\theta}{\beta^{d-1}_S(x,\theta)}\,,
\eea
where once again the factor of $4$ comes from the separation of the integral into four regions which contributes the same to $g_S(A)$. This integral can be computed analytically for individual integer values of $d$ from which one can derive the general formula to be 
\bea\label{g-strip-result}
g_S(A)=-\frac{\Gamma\(\frac{d-1}2\)}{\Gamma\(\frac d2\) \pi^{d-\frac{1}{2}}}\frac{A_\perp}{l^{d-2}}\log\(\frac{2\epsilon}{l+\sqrt{l^2-4\epsilon^2}}\)+{\cal O}(\epsilon) \,,
\eea
up to terms that go to zero as $\epsilon$ goes to zero.

Adding the results of both contributing regions (\ref{g-rindler-result}) and (\ref{g-strip-result}) we find 
 \bea
g(A)=\frac{2}{(d-2)(2\pi)^{d-1}}\frac{A_\perp}{\epsilon^{d-2}}-\frac{\Gamma\(\frac{d-1}2\)}{(d-2)\Gamma\(\frac d2\) \pi^{d-\frac{1}{2}}}\frac{A_\perp}{l^{d-2}}\,.
\eea
From our conjectured formula we can thus evaluate the Rényi-$0$ for a free scalar and a free fermion using the associated Stefan-Boltzmann constants 
\bea
S_n(A)=\frac{\sigma_{B/F}}{n^{d-1}} \left[\frac{2}{(d-2)(2\pi)^{d-1}}\frac{A_\perp}{\epsilon^{d-2}}-\frac{\Gamma\(\frac{d-1}2\)}{(d-2)\Gamma\(\frac d2\) \pi^{d-\frac{1}{2}}}\frac{A_\perp}{l^{d-2}}\right]\,.
\eea
From this formula, it is easy to compute the constants defined in the previous subsection which were computed using exact methods
\bea
\kappa^{s/f}_n(d)=\frac{\sigma_{B/F}}{n^{d-1}}\frac{\Gamma\(\frac{d-1}2\)}{(d-2)\Gamma\(\frac d2\) \pi^{d-\frac{1}{2}}}\,.
\eea
These coefficients agree exactly with (\ref{kappa-strip-s}) and (\ref{kappa-strip-f}) after replacing the values of $\sigma_{B/F}$ for the scalar (\ref{sigmaB}) and the fermion (\ref{sigmaF}) respectively.

\section{State propagation and entropy current}

In this section, we study the characterization of the state at large modular temperatures in terms of their ET. This is a high-temperature thermal ensemble described as a gas of relativistic (massless) particles with thermal distribution.
The description is in terms of a free relativistic Boltzmann equation, that is reviewed below. The Boltzmann distribution function $\psi(x,p)$ contains the information on the ET. The statistical system under study can be alternatively described by an infinite tower of conserved currents, whose conservation follows from the ballistic propagation of the ET described in Sec. \ref{Lorentz-ET}. When a conformal Killing vector is present, we obtain this tower of conserved currents explicitly. The example of the strip geometry in which there is no Killing vector, but the ET are known, is also presented.

\subsection{Free relativistic Boltzmann equation}

Consider a general QFT in $d$ dimensions with a free UV fixed point on a manifold with metric $g_{\mu\nu}$. In this setup, the high energy physics is described by ultra-relativistic collisionless particles that travel at the speed of light with adequate spin statistics corresponding to the fundamental UV fields. 
These particles travel along null geodesics with momenta $p^\mu$, with $p^2=0$.
One can define a Lorentz invariant distribution function $\psi(x,p)$ for the number of particles $dN_\Sigma$ that traverse a $d-1$ volume element $\Sigma$ with momentum $p^\mu$. The surface $\Sigma$ itself is defined as normal to an observer with a world-line tangent to the unitary time-like vector $\hat{n}^\mu$. The energy of a particle of momentum $p^\mu$ measured by this observer is $p^\mu \hat{n}_\mu$.

In the literature, see e.g. \cite{walker_1936,BoltzmannKremer:2002,BoltzmannCovariant} there is more than one presentation of $\psi$ depending on which variables are thought of as fundamental. The standard presentation is to take a function of $x^\mu$ and $p^i$, i.e. only the space-like momenta, where the on-shell constraint on $p^2$ is already taken into account and $p^0$ must be regarded as a function of the other variables $p^0=p^0(x,p^i)$. There is also another presentation using $x^\mu$ and $p_i$, i.e. using the dual space-like momenta.
In this work, we will be more interested in a manifestly covariant presentation of $\psi$. This is, we define $\psi(x,p)$ as an off-shell function of the complete contravariant vectors $\{x^\mu,p^\nu\}$ but whose physical information is extracted upon projecting on-shell, i.e. $p^2=0$ in our scenario.

Our Boltzmann function $\psi(x,p)$ is thus defined by the relation
\begin{equation}\label{psi-definition}
    dN_\Sigma \equiv \psi(x,p) \; (\hat{n}\cdot p)\;  d\Sigma \; d\Pi \;.
\end{equation}
where we have defined the covariant volume elements
\begin{equation}
    d\Sigma \equiv \sqrt{-g_\Sigma} \; \prod_{i=1}^{d-1}d x^i \;,\qquad\qquad
    d\Pi \equiv \sqrt{-g}\, \delta(p^2)\;2\,\theta(\hat{n}\cdot p) \frac{\prod_{\alpha=0}^{d-1}\,d p^\alpha}{(2\pi)^{d-1}}\;,
\end{equation}
where $\sqrt{-g_\Sigma}$ is the induced metric on $\Sigma$. Since $dN_\Sigma$, $d\Sigma$, $d\Pi$, and $(\hat{n}_\mu p^\mu)$ are Lorentz invariant, we have that $\psi(x,p)$ is itself a Lorentz invariant probability density. This is also independent of the observed direction $\hat{n}$ because $d\Sigma\, (\hat{n}\cdot p)$ is independent of $\hat{n}$.

The Boltzmann equation is derived by demanding $dN_\Sigma$ to be invariant under an infinitesimal evolution on the proper time $\lambda$ of all particles. Then, the geodesic equations
\begin{equation}\label{GeodTrans0}
    \frac{dx^\mu}{d\lambda} \equiv p^\mu \;,\qquad \qquad \frac{d^2 x^\mu}{d\lambda^2} + \Gamma^\mu_{\alpha\beta} \frac{dx^\alpha}{d\lambda}\frac{dx^\beta}{d\lambda}= \frac{dp^\mu}{d\lambda} + \Gamma^\mu_{\alpha\beta}\;p^\alpha p^\beta=p^\alpha \nabla_\alpha p^\mu=0 \;,
\end{equation}
impose non linear transformations for space and momenta,
 \begin{equation}\label{GeodTrans}
   dx^\mu = p^\mu d\lambda \;,\qquad \qquad d p^\mu =- \Gamma^\mu_{\alpha\beta}\;p^\alpha p^\beta d\lambda \;.
\end{equation}
One can explicitly check that $d\Pi$ and $\big(d\Sigma\, (\hat{n}\cdot p)\big)$ are invariant on their own under these transformations \cite{BoltzmannKremer:2002,BoltzmannCovariant}, so that we must impose $\psi(x,p)$ to be invariant under \eqref{GeodTrans}. To do this, one must regard $\psi(x,p)$ as a function only on the proper time $\lambda$ and demand 
\begin{equation}\label{Boltzmann-eq}
\frac{d\psi\big(x(\lambda),p(\lambda)\big)}{d\lambda}
    = \frac{\partial \psi(x,p)}{\partial x^\mu}p^\mu- \frac{\partial \psi(x,p)}{\partial p^\mu}\Gamma^\mu_{\alpha\beta} \,p^\alpha\, p^\beta=0\,,
\end{equation}
where \eqref{GeodTrans} was used. The last equation is the covariant version of the Boltzmann equation on curved manifolds, which imposes an evolution equation for the invariant phase space density function. This should be complemented with the on-shell constraint, i.e. $p^2=0$.

For the state in the limit, $n\to 0$ the density of states is the occupation number corresponding to the  ET  
\begin{equation}\label{Srel-cov}
    \psi(x,p)  \equiv \frac{1}{ e^{n S_{\rm rel}(x,p) }\mp 1}\;.
\end{equation}
From \eqref{eqq}, we know that the relative entropy, in our regime of interest, is a homogeneous function of the momentum variable, this is, for $\lambda>0$, we have $S_{\rm rel}(x,\lambda\, p)= \lambda\, S_{\rm rel}(x, p)$. Therefore, choosing $\lambda=-(\hat n\cdot p)^{-1}$, with $\hat n$ an arbitrary future oriented time-like unit vector, we get
\bea
S_{\rm rel}(x, p)=-(\hat n \cdot p ) \, S_{\rm rel}\(x, -\frac{p}{(\hat n \cdot p )}\)\equiv -(\hat n \cdot p )\,\beta(x,\hat{p}), \quad {\rm where}\quad \hat{p}^\mu\equiv-\frac{p^\mu}{(\hat{n}\cdot p)}
\eea
This equation provides the proper generalization of the notion of local temperatures \eqref{eqq}. The above relation is understood on-shell, but it is harmlessly extended off-shell in an arbitrary way just for simplicity of the notation. 

Now, one can see that eqs. \eqref{Boltzmann-eq} and \eqref{Srel-cov} imply that $S_{\rm rel}(x,p)$ itself has to meet the Boltzmann equation \eqref{Boltzmann-eq}. 
 This in fact follows from our discussion in Sec. \ref{Lorentz-ET} using the analytic extension of the Euclidean eikonal analysis of \cite{Arias:2016nip,arias2017anisotropic}. One can see that eqs. \eqref{L-EikoEqs} and \eqref{rela} also impose a differential equation for $S_{\rm rel}(x,p)$ under geodesic transport \eqref{GeodTrans0}. We now show that these equations lead to the Boltzmann equation for $S_{\rm rel}(x,p)$. By extending all quantities off-shell we get under the flow of particles 
\begin{equation}
   T^{\mu} = \frac{p^\mu}{S_{\rm rel}(x,p)} \qquad\Rightarrow\qquad T^\alpha \nabla_\alpha  T^\mu = \frac{1 }{S_{\rm rel}}p^\alpha\nabla_\alpha p^\mu -
\frac{p^\mu }{S_{\rm rel}^2}\;  p^\alpha \nabla_\alpha S_{\rm rel}(x,p)=0\;.
\end{equation}
By using eq. \eqref{GeodTrans0} one can see that the first term vanishes. The second term can also be manipulated using the geodesic equation to obtain,
\begin{equation}
    p^\alpha \nabla_\alpha S_{\rm rel}(x,p)=
    p^\alpha \partial_\alpha S_{\rm rel}(x,p) = \frac{d x^\alpha}{d\lambda}\frac{ d S_{\rm rel}(x,p)}{d x^\alpha}=\frac{d S_{\rm rel}(x,p)}{d \lambda}=0\;,
\end{equation}
which is exactly \eqref{Boltzmann-eq}. More generally, any scalar function of $S_{\rm rel}(x,p)$ will also meet \eqref{Boltzmann-eq}. We emphasize that once written as in the right-hand side of \eqref{Boltzmann-eq}, the Boltzmann equation holds as a partial differential equation on $\{x^\mu,p^\nu\}$ regardless of the geodesic motion \eqref{GeodTrans0} we used to derive it.

\subsection{A tower of conserved currents}

The Boltzmann equation \eqref{Boltzmann-eq} in turn implies the conservation of an infinite number of currents \cite{BoltzmannKremer:2002}. Consider for example the particle number current, defined as 
\bea\label{J-Number}
N^\mu(x)\equiv\int d\Pi\,  p^\mu\,\psi(x,p) =\int \frac{d^dp}{\(2\pi \)^{d-1}} \,\sqrt{-g}\, 2\theta(p^0)\, \delta(p^2)\, p^\mu\, \psi(x,p)\,.
\eea
We now want to show that this current is conserved. We begin by taking an ordinary spatial derivative. Notice first that the momentum and coordinate variables are independent of each other\footnote{This is not necessary, if one considers the deformation of the tangent space under an infinitesimal displacement in \eqref{J-Number} one can show that these contributions cancel by themselves \cite{walker_1936}, in agreement with our statement.}, so that \eqref{GeodTrans0} does not apply, but the partial derivative version of \eqref{Boltzmann-eq} does. We have
\begin{align}\label{Nmu-cons}
\partial_\mu N^\mu&=\int \frac{d^dp}{\(2\pi \)^{d-1}}\, 2\theta(p^0)\,\sqrt{-g}\,\[\frac{1}{\sqrt{-g}}\frac{\partial \sqrt{-g}}{\partial x^\mu} \delta(p^2) + 
 \delta'(p^2)
\frac{\partial g_{\alpha \beta}}{\partial x^\mu}p^\alpha p^\beta \]p^\mu \psi(x,p) \nonumber \\
& \qquad +\int \frac{d^dp}{\(2\pi \)^{d-1}}\,\sqrt{-g}\,2\theta(p^0)\, \delta(p^2)\, p^\mu \frac{\partial \psi(x,p)}{\partial x^\mu}\,.
\end{align}
In order to connect to the Boltzmann equation, we would like to find the term
\bea
\Gamma^{\mu}_{\alpha \beta} p^\alpha p^\beta \delta(p^2)\frac{\partial \psi(x,p)}{\partial p^\mu}\,.
\eea
We do so by adding a total momenta derivative of the form
\begin{align}\label{total-deriv}
\frac{\partial}{\partial p^\mu}\[ \Gamma^{\mu}_{\alpha \beta} \,p^\alpha p^\beta \delta(p^2) \psi(x,p)\]&=2\, \Gamma^{\mu}_{ \mu \alpha}\, p^\alpha \delta(p^2) \psi(x,p)+\Gamma^{\mu}_{\alpha \beta} p^\alpha p^\beta \frac{\partial \delta(p^2)}{\partial p^\mu} \psi(x,p)\nonumber \\
&\qquad+\Gamma^{\mu}_{\alpha \beta} p^\alpha p^\beta \delta(p^2)\frac{\partial \psi(x,p)}{\partial p^\mu}\,.
\end{align}
Interestingly, the second term in the integrand of (\ref{Nmu-cons}) coincides exactly with the second term of the right-hand side of (\ref{total-deriv}) after using the following identities 
\bea
\frac{\partial \delta (p^2)}{\partial p^\mu}=2 \delta'(p^2-m^2) p_\mu\,, \qquad \qquad p^\alpha p^\beta p^\mu \frac{\partial g_{\alpha \beta}}{\partial x^\mu}=2\,\Gamma^{\mu}_{\alpha \beta} p^\alpha p^\beta p_\mu \,.
\eea
In this way equation (\ref{Nmu-cons}) reduces to
\begin{align}\label{Nmu-cons-2}
\partial_\mu N^\mu&=\int \frac{d^dp}{\(2\pi \)^{d-1}} \,\sqrt{-g}\,2\theta(p^0)\,\delta(p^2)\,\[ \( p^\mu \frac{\partial \psi(x,p)}{\partial x^\mu}-\Gamma^{\mu}_{\alpha \beta} p^\alpha p^\beta \frac{\partial \psi(x,p)}{\partial p^\mu}\) -\Gamma^{\alpha}_{\alpha \mu} p^\mu  \psi(x,p)\] \nonumber \\
& \qquad  +\int \frac{d^dp}{\(2\pi \)^{d-1}} \,\sqrt{-g}\,2\theta(p^0) 
\frac{\partial}{\partial p^\mu}\[ \Gamma^{\mu}_{\alpha \beta} \,p^\alpha p^\beta \delta(p^2) \psi(x,p)\]\,,
\end{align}
where we also use the well-known identity $
\partial_\mu \sqrt{-g}=\sqrt{-g}\;\Gamma^{\alpha}_{\alpha \mu}$. The integral in the second line leads to a boundary term that vanishes by the assumption that $\psi(x,p)$ decays faster than any power of the momentum. By completing the covariant divergence of $N^\mu$ using the last term in the first line above, we find that the Boltzmann equation implies the conservation of $N^\mu$:
\bea
\nabla_\mu N^\mu=\int \frac{d^dp}{\(2\pi \)^{d-1}} \,\sqrt{-g}\,2 \theta(p^0)\delta(p^2)\,\[  p^\mu \frac{\partial \psi(x,p)}{\partial x^\mu}-\Gamma^{\mu}_{\alpha \beta} p^\alpha p^\beta \frac{\partial \psi(x,p)}{\partial p^\mu}\]=0\,,
\eea
as we wanted to prove. 

In the same way one can build an infinite tower of conserved currents of the form
\begin{equation}\label{Tmunu-Currents}
     T^{\mu\nu}=\int d\Pi \;  p^{\mu} \;p^{\nu}\; \psi(x,p) \;,\qquad\qquad F^{\mu_1\dots\mu_m}(x) \equiv \int d\Pi \;\left( \prod_{ i=1}^m p^{\mu_i}\right)\; \psi(x,p) \;,
\end{equation}
where we have singled out the stress-energy tensor of the system. All these currents can be shown to be conserved in any of its indices in an identical fashion as with $N^{\mu}$ by virtue of the Boltzmann equation \eqref{Boltzmann-eq}. Furthermore, they are symmetric in all of their indices by construction and traceless under any contraction. 

 \subsection{Conserved currents in terms of the ET \label{conserved-currents}}

Consider a general current in a scenario where $\psi(x,p)=\psi\[n\, \beta(x,\hat p)\,(-\hat{n}\cdot p)\]$, where we remind the reader that $\hat{p}=p/(-\hat{n}\cdot p)$. We have
\begin{align}
F^{\mu_1\dots\mu_m}(x) 
&= \int d\Pi \;\left( \prod_{ i=1}^m p^{\mu_i}\right)\; \psi\[n \, \beta(x,\hat p)\,(-\hat{n}\cdot p)\] \nn\;.
\end{align}
Now, let us consider a local change of coordinates $x=f(\tilde{x})$ which takes $g_{\mu \nu}(x)$  at $x=q$ into $\eta_{\mu \nu}$ at $\tilde{x}=f^{-1}(q)$, namely
\bea\label{coor-gtoeta}
g_{\mu \nu}(x)\frac{\partial x^\mu}{\partial \tilde{x}^\alpha} \frac{\partial x^\nu}{\partial \tilde{x}^\beta}\Big|_{x=q}=\eta_{\alpha \beta}\,.
\eea
This change of coordinates will induce a corresponding change of coordinates in the momentum from $p^\mu$ to $\tilde{p}^\nu$ where the latter lives in flat space
\bea
p^\mu=\frac{\partial x^\mu}{\partial \tilde{x}^\nu}\tilde{p}^\nu\;.
\eea
This results in 
\begin{align}
F^{\mu_1\dots\mu_m}(x) 
&= \prod_{ i=1}^m \(\frac{\partial x^{\mu_i}}{\partial \tilde{x}^\nu_i}\) \int d\tilde{\Pi} \;\left( \prod_{ i=1}^m \tilde{p}^{\nu_i}\right)\; \psi\[n \, \beta\(x,\(\frac{\partial x^\mu}{\partial \tilde{x}^\nu}\)\frac{\tilde{p}^\nu}{(-\hat{\tilde{n}}\cdot \tilde{p})}\)\,(-\hat{\tilde{n}}\cdot \tilde{p})\]\,. \nn
\end{align}
The momentum integral in the tilde coordinates is in flat space, and thus we can use spherical coordinates to carry out the integral on $|\vec{\tilde{p}}\,|$. Notice that $\tilde{p}^0=\tilde{p}_0=|\vec{\tilde{p}}\,|$, thus, if we chose $\hat{\tilde{n}}^\mu=\delta^{\mu}_{0}$, we have $\hat{\tilde{p}}^\mu=\tilde{p}^\mu/\tilde{p}^0$ and
\begin{align}
F^{\mu_1\dots\mu_m}(x) 
&= \prod_{ i=1}^m \(\frac{\partial x^{\mu_i}}{\partial \tilde{x}^\nu_i}\)\int \!\frac{d\tilde{\Omega}_{d-2}}{(2\pi)^{d-1}}\int_0^\infty \hspace{-.3cm}d\,|\vec{\tilde{p}}\,||\vec{\tilde{p}}\,|^{d-3+m}\left( \prod_{ i=1}^m \hat {\tilde{p}}^{\nu_i}\right) \psi\[n \,|\vec{\tilde{p}}\,| \, \beta\(x, \(\frac{\partial x^\mu}{\partial \tilde{x}^\nu}\)\hat{\tilde{p}}^\nu\)\]  \nn \\
&= \prod_{ i=1}^m \(\frac{\partial x^{\mu_i}}{\partial \tilde{x}^\nu_i}\)  \hspace{-.1cm} \frac{1}{n^{d-2+m}}  \hspace{-.1cm}\int \hspace{-.1cm} \frac{d\tilde{\Omega}_{d-2}}{(2\pi)^{d-1}} \frac{ \prod_{ i=1}^m \hat{\tilde{p}}^{\nu_i}}{\beta\(x, \(\frac{\partial x^\mu}{\partial \tilde{x}^\nu}\)\hat{\tilde{p}}^\nu\)^{d-2+m}}\left(\int_0^\infty \hspace{-.3cm} dl \; l^{d-3+m}\; \psi[l] \right)\nn\\
&\equiv \prod_{ i=1}^m \(\frac{\partial x^{\mu_i}}{\partial \tilde{x}^\nu_i}\) \frac{f(m)}{n^{d-2+m}}\int d\tilde{\Omega}_{d-2} \frac{ \prod_{ i=1}^m \hat{\tilde{p}}^{\nu_i}}{\beta\(x, \(\frac{\partial x^\mu}{\partial \tilde{x}^\nu}\)\hat{\tilde{p}}^\nu\)^{d-2+m}}\,,\;
\end{align}
where $ l=n \,|\vec{\tilde{p}}\,| \,  \beta\(x,\(\frac{\partial x^\mu}{\partial \tilde{x}^\nu}\)\hat{\tilde{p}}^\nu\) $
and $f(m)$ are ``chemical'' coefficients that depend on the nature of the microscopic components of the system. For example, for the cases of bosons and fermions one gets,
\begin{equation}
    f_b(m)\equiv (2\pi)^{1-d} \zeta (d-2+m)  \Gamma (d-2+m)\;,
   \qquad\qquad f_f(m)=(1-2^{3-d-m})f_b(m)\;.
\end{equation}
In general, these factors are non-trivially related to the Stefan-Boltzmann constants except for $m=2$, c.f. \eqref{sigmaB} and \eqref{sigmaF}
\begin{equation}
    f(2)= \frac{ (d-1)}{{\rm vol}{(\mathbb{S}^{d-2})}}\;\sigma\;,
\end{equation}
which is the correct normalization for $T_{\mu\nu}$ to yield the energy of the gas.

Finally, we can undo the coordinate transformation and express everything in terms of the original coordinates and variables as
\bea
F^{\mu_1\dots\mu_m}(x) = \frac{f(m)}{n^{d-2+m}}\int d\tilde{\Omega}_{d-2} \frac{ \prod_{ i=1}^m \hat{p}^{\nu_i}}{\beta(x, \hat{p}^\nu)^{d-2+m}}\,,\label{Fmunu-Angles}
\eea
where 
\bea
\quad \hat{p}^\mu=\(\frac{\partial x^\mu}{\partial \tilde{x}^\nu}\)\hat{\tilde{p}}^\nu \;,\qquad{\rm and} \qquad g_{\mu \nu}(x)\frac{\partial x^\mu}{\partial \tilde{x}^\alpha} \frac{\partial x^\nu}{\partial \tilde{x}^\beta}\Big|_{x=q}=\eta_{\alpha \beta}\,,\label{hatp-hatptilde}
\eea
where $\hat{\tilde{p}}^0=\tilde{p}^0/(- \tilde n \cdot \tilde p)=1$. In (\ref{Fmunu-Angles}), one must keep in mind that the angular integral is characterizing the direction of the flat space momentum variables, $\tilde{p}^\mu$.
A key conclusion of this analysis, see \eqref{Fmunu-Angles} and \eqref{hatp-hatptilde}, is that the conserved currents we define contain the information of all moments of $\beta(x,\hat p)^{-1}$ distribution and thus knowing all conserved currents in spacetime formally contains the same information as $\beta(x,\hat p)$. 

A comment on covariance and normalization is due. Notice first that the conservation of the currents is independent of its normalization. In particular, $N^\mu$ in eq. \eqref{J-Number} and $T^{\mu\nu}$ in \eqref{Tmunu-Currents} are normalized to give the particle number and energy density at each point respectively. On the other hand, notice also that we are dealing with a single quantity with units, $\beta(x,\hat p)$, so that there is also a single covariant quantity that can be defined for any given dimension, essentially
\begin{align}\label{gral-currents}
\tilde F^{\mu_1\dots\mu_m}(x) \sim \int d\tilde{\Omega}_{d-2} \frac{ \prod_{ i=1}^m \hat p^{\mu_i}}{\beta(x,\hat p)^{d-2+m}}\;,
\end{align}
with $\hat{p}^\mu$ defined in \eqref{hatp-hatptilde}.
Above, we have built for $m=1$ in eq. \eqref{J-Number} the particle number density $N^\mu$. However, to make contact with our analysis in Sec. \ref{Sec:Renyi-O} we would rather work with another density normalized to provide the Rényi-$0$ in eq. \eqref{Rényi-zero}. The current of interest is
\begin{equation}
    J^\mu \equiv \frac{ \sigma}{n^{d-1}} \frac{1}{{\rm vol}{(\mathbb{S}^{d-2})}}\int d\tilde{\Omega}_{d-2} \frac{ \hat p^{\mu}}{\beta(x,\hat p)^{d-1}} \qquad\Rightarrow\qquad  S_n(\Sigma)=\int_\Sigma d\Sigma \; \hat n_\mu\;J^\mu(x)\;,
\end{equation}
so that for unitary time-like $\hat n$ we recover \eqref{Rényi-zero-formula}. The above equation for general metric are defined according to \eqref{hatp-hatptilde}.

We conclude that the present analysis gives an infinite tower of symmetric traceless conserved currents \eqref{Fmunu-Angles} that contains the same information as the $\beta$'s. In what follows we compute explicitly some of these currents.

\subsection{Examples}

In this subsection, we would like to construct some explicit realizations of the conserved currents $J^\mu$ and $T^{\mu \nu}$ discussed in the previous subsections.  

\subsubsection{Multi-intervals}

The local temperatures at $t=0$ for QFTs in $d=2$ for $N$ intervals were reviewed in Sec. \ref{Sec3Multi-interval} and are
\bea
\beta(x)
=2\pi\(\sum_{i=1}^N \[\frac{1}{x-l_i}+\frac{1}{r_i-x}\] \)^{-1}\;,
\eea
where there is no momentum parameter since the two possible directions have the same ET at $t=0$. The extension to $t\neq 0$ in $d=2$ is particularly easy as the $d\Omega_{d-2}$ collapses into a sum over the two rays $x \pm t$, i.e. $\beta_\pm(x,t) \equiv \beta(x\mp t)$ and ${\rm vol}{(\mathbb{S}^{0})}\equiv2$, see footnote \ref{footd2}. Thus the current can be computed as
\bea
J_\mu(x,t) = \frac {\sigma}{ 2n} \left( \frac{p^+_\mu}{\beta_+(x,t)}+\frac{p^-_\mu}{\beta_-(x,t)}\right)=\frac{\sigma\, u^\mu}{n \beta(x,t)}\;,
\eea
where $p^\pm_\mu=\{ 1,\pm 1\}$,
\be
\beta(x,t)=\sqrt{\beta_+(x,t) \beta_-(x,t)}\,,
\ee
and we have defined a velocity $u^\mu$, $u^2=-1$, given by
\bea
u^\mu = \frac{\hat{t}^\mu}2\[\sqrt{\frac{\beta_-(x,t)}{\beta_+(x,t)}}+\sqrt{\frac{\beta_+(x,t)}{\beta_-(x,t)}}\]+\frac{\hat{x}^\mu}2\[\sqrt{\frac{\beta_-(x,t)}{\beta_+(x,t)}}-\sqrt{\frac{\beta_+(x,t)}{\beta_-(x,t)}}\].
\eea

 The stress-energy tensor can also be built in a similar fashion, yielding
\bea
T_{\mu\nu}(x,t) = \frac {\sigma}{ 2 n ^2} \left( \frac{p^+_\mu p^+_\nu}{\beta_+(x,t)^2}+\frac{p^-_\mu p^-_\nu}{\beta_-(x,t)^2}\right)=\frac{P }{n^2}\(2 u^\mu u^\nu +\eta^{\mu \nu}\)\;,
\eea
where $P=\sigma /\beta(x,t)^{2}$. 
Notice that the system has the current and stress tensor of a perfect fluid with fluid velocity $u^\mu$, pressure $P$, and energy density $\rho=P$.
We will see that the perfect fluid structure only holds in cases where a Killing vector is present. 
\subsubsection{Rindler}

For Rindler, the local temperatures on the $t=0$ slice are given by
\bea
\beta(x, \hat{p})=2\pi x \,,
\eea
where $x$ is the coordinate orthogonal to the boundary. For concreteness, we evaluate the currents for $d=3$, since the general $d$ result is a straightforward generalization. 

Given the initial data at $t=0$, the Lorentizian eikonal equations (\ref{L-EikoEqs}) tell us how to propagate that data to $t\geq 0$. Given a point $(t,x)$ with $0<t<x$, and a direction $\hat{p}$, the local temperature $\beta(x,t,\hat{p})$ can be obtained by propagating backward this data to the $t=0$ surface, namely
\bea\label{beta-rindler}
\beta(x,t,\hat{p})=
2\pi( x-t\cos\theta)\,,
\eea
where the direction of $\hat{p}(\theta)=\{1, \cos\theta, \sin\theta\}$ is given by the angle $\theta$ between the momentum spacial direction $\hat{p}$ and $x$. We find
\bea\nn
J^{\mu} =\frac{\sigma}{2\pi n^2} \, \int_0^{2\pi} \hspace{-.25cm} d \theta \frac{\hat{p}^\mu }{\beta(x,t,\hat{p})^2} = \frac{\sigma}{(n \beta_R)^2}\( \hat{x}\frac{t}{\sqrt{x^2-t^2}} +\hat{t}\frac{x}{\sqrt{x^2-t^2}} \)^\mu \equiv \frac{\sigma}{(n \beta_R)^2}  u^{\mu}\;,
\eea
with $\beta_R \equiv 2\pi \sqrt{x^2-t^2}$, $ u^\mu$ a velocity, and
\bea
T^{\mu\nu} =\frac{\sigma}{\pi n^{3}}\int_0^{2\pi} \hspace{-.25cm} d\theta\frac{\hat p^{\mu}\hat p^{\nu}}{\beta(x,t,\hat{p})^3}=\frac{\sigma}{(n \beta_R)^3}(3\,   u^\mu  u^\nu+\eta^{\mu \nu})\;.
\eea
For general dimensions, the expressions can be readily generalized to
\bea\label{current-Killing}
J^\mu=  \frac{\sigma}{(n \beta_R)^{d-1}} u^\mu\,,
\qquad 
T^{\mu \nu}=  \frac{\sigma}{(n \beta_R)^{d-2}}\,\( d\, u^\mu u^\nu +\eta^{\mu \nu}\) \,,
\qquad 
(d-1)J^\mu=u_\mu T^{\mu \nu}\,,
\eea
where $u^\mu$ is completed with trivial entries in all dimensions other than $\{t,x\}$.

\subsubsection{ET given by a conformal Killing vector}\label{KillingCurrents}

The previous examples are all cases where the ET can be defined in terms of a conformal Killing vector, $\xi$. As reviewed in section \ref{Lorentz-ET}, in such cases, the modular Hamiltonian adopts the simple local form (\ref{local}) which leads to the closed formula (\ref{236}) for the temperature vector $T^\mu$. From these results, the general formulas for the conserved currents (\ref{Fmunu-Angles}) can be integrated exactly. For instance, for $J^\mu$ and $T^{\mu \nu}$ we have  
\bea\label{J-Killing}
J^\mu=\frac{\sigma}{n^{d-1}}\frac{1}{{\rm vol}{(\mathbb{S}^{d-2})}}\int d\tilde{\Omega}_{d-2} \frac{   \hat p^{\mu}}{\( -\hat{p}\cdot \xi \)^{d-1}}= \frac{\sigma}{n^{d-1}}\frac{\xi^{\mu}}{\(-\xi^2\)^{\frac d2}}\,,
\eea
\bea\label{T-Killing}
T^{\mu \nu}=\frac{\sigma}{n^{d}}\frac{(d-1)}{{\rm vol}{(\mathbb{S}^{d-2})}}\int d\tilde{\Omega}_{d-2} \frac{  \hat p^{\mu} \hat p^{\nu}}{\( -\hat{p}\cdot \xi \)^{d}}= \frac{\sigma}{n^{d}}\frac{d}{\(-\xi^2\)^{\frac{d+2}2}}\(\xi^\mu \xi^\nu -\frac{1}{d} g^{\mu \nu} \xi^2 \)\,,
\eea
where we used the relation $\beta(x,\hat{p})=-\hat{p}\cdot \xi$ which follows from (\ref{236}), (\ref{esaab}) and (\ref{KA-local}). This pair of current energy-momentum tensor has the form of a perfect fluid with velocity vector $u^\mu$ and density $\rho=P$ given by 
\bea
u^\mu=\frac{\xi^\mu}{\(-\xi^2\)^{\frac12}}, \qquad \rho=\frac{\sigma}{\(-\xi^2\)^{\frac{d}{2}}}\,.
\eea
Similar formulas can be obtained for the higher index generalizations (\ref{Fmunu-Angles}), which can be alternatively derived from general considerations. For instance (\ref{J-Killing}), (\ref{T-Killing}) are the only vector and two-index traceless symmetric tensor that can be built out of $\xi^\mu$ and $\eta^{\mu \nu}$ with the appropriate scaling dimension up to an overall dimensionless factor. For the three-index tensor following this prescription one gets
\bea
F^{\mu \nu \rho}= \frac{c_3}{n^{d+1}} \frac{1}{\(-\xi^2\)^{\frac{d+4}2}}\[ \xi^{\mu}\xi^{\nu}\xi^{\rho}-\frac1{d+2}\( \xi^\mu g^{\nu \rho}+\xi^\nu g^{\mu \rho}+\xi^\rho g^{\mu \nu}\) \xi^2 \]\,.
\eea
where the coefficient $c_3$ can be fixed from (\ref{Fmunu-Angles}) to be
\bea
c_3=f(3)\(\frac{d+2}{d-1}\) {\rm vol}{(\mathbb{S}^{d-2})}\,.
\eea
The general formula (\ref{Fmunu-Angles}), which in this context takes the form 
\bea\label{gen-J}
J_{m}^{\mu_1\cdots \mu_{m}}=\frac{f(m)}{n^{d-2+m}} \int d\tilde{\Omega}_{d-2} \frac{ \hat p^{\mu_1}\cdots \hat p^{\mu_m}}{\(-\hat p \cdot \xi\)^{d-2+m}}\,,
\eea
provides useful relations between the currents
. For instance, 
\bea\label{low-from-high}
\xi_{\mu_m}J_m^{\mu_1\cdots \mu_{m}}=-\frac{1}{n}\frac{f(m)}{f(m-1)}J_{m-1}^{\mu_1\cdots \mu_{m-1}}
\eea
follows easily from it. This equation gives a recursive way to obtain all the lower index currents from the higher ones.  Another interesting relation  
\bea
\nabla^{\mu_1}J_{m-1}^{\mu_2 \cdots \mu_n}=\frac{n \,(d+m-3)}{d} \frac{f(m-1)}{f(m)}\(\nabla \cdot \xi\) J_m^{\mu_1\mu_2 \cdots \mu_n}
\eea
connects the covariant derivative of $J_{m-1}$ and  $J_m$. This equation follows from (\ref{gen-J}) and (\ref{conf-Killing}).  
Contracting indices, one gets a relation between the conservation of $J_{m-1}$ and the zero trace condition of $J_{m}$. Finally, one can  construct higher index currents from the lower ones, this is the converse of (\ref{low-from-high}) as 
\bea\label{J-recursive-Killing}
J_m^{\mu_1 \cdots \mu_{m-1} \mu_m}=\frac{1}{n}\frac{c_m}{\(-\xi^2 \)}\[ J_{m-1}^{(\mu_1 \cdots \mu_{m-1}}\xi^{\mu_m)}
-\kappa_m\, \xi_\sigma J_{m-1}^{\sigma ( \mu_1 \cdots \mu_{m-2}}g^{\mu_{m-1} \mu_{m})}\]\,,
\eea
where $\kappa_m$ is determined from the zero trace condition, and $c_m$ from (\ref{gen-J})  
\bea
\kappa_m=\frac{m-1}{d+2(m-2)}\,, \quad c_m= \frac{d+2(m-2)}{d+m-3}\(\frac{f(m)}{f(m-1)}\)\,.
\eea
Formula (\ref{J-recursive-Killing}) provides an explicit construction of all the associated conserved currents.  

\subsubsection{Strip}

We now present a final example of this formalism in which a conformal Killing vector is not present. The ET for the strip were found in \cite{arias2017anisotropic} for any dimension. Here we present a simplified analytic expression for the ET, we repeat here \eqref{beta-strip2} for reference,
\begin{equation*}
\beta(x,\hat p)\equiv
\begin{cases}
\pi(1-2 |x|)& \text{ for }  \sin\theta > 1-2|x|\\
\frac\pi2 \frac{\cos^2\theta-4x^2}{1-|\sin\theta|} & \text{ for }  \sin\theta<1-2|x|
\end{cases}\;.
\end{equation*}
These are the ET at $t=0$ for a strip whose walls sit at $x=\pm1/2$. Notice the $|\sin\theta|$ in the denominator. Notice also that $1-|\sin\theta|\neq0$ is the domain where the expression is valid.
We will work once again with $d=3$, where a single angle $\theta\in[0,2\pi)$ parametrizes the ET. We will present results for the currents at $t=0$. This will already reveal that not all systems described in this fashion meet the structure of an ideal fluid. A complete analytic expression for $t\neq0$ seems to be out of reach.

For $J_\mu$ in $d=3$ we get 
\begin{align}
    J^{\mu}(x,t=0) 
        = \frac{\sigma}{2\pi n^{2}}\int_{0}^{2\pi} \hspace{-.25cm} d\theta \;\frac{\hat p^{\mu}}{\beta(x,\hat p)^2 }\;,
\end{align}
which for $0<x<1/2$ and $t=0$ yields, 
\begin{align}\nn
    J^{t}(x,0)&=\frac{\sigma}{\pi^3 x^3 n^{2}} \left(\frac{2 x^3 \cos^{-1}(1-2 x)}{ (1-2 x)^2} + \frac{2x \left(1-(1+x)\sqrt{(1-x) x} \right)}{ \left(1-4
   x^2\right)} -\tanh ^{-1}(2
   x) \right.\\
   &\qquad\qquad\qquad\left.+\tanh
   ^{-1}\left(\frac{\sqrt{x}}{\sqrt{1-x}}\right)-\frac{\left(1-6 x^2\right) }{\left(1-4
   x^2\right)^{3/2}}\tanh ^{-1}\left(\frac{\sqrt{x}
   (1-2 x)}{\sqrt{1-x} \sqrt{1-4 x^2}}\right)\right),\nn
\end{align}
\begin{equation}
    J^{x}(x,t=0)=J^{y}(x,t=0)=0\;.
\end{equation}
The solution is symmetric in $x\to-x$. This analytic expression does not hold beyond $|x|>1/2$. The spatial pieces $J^{x}(x,t=0)$ and $J^{y}(x,t=0)$ are zero by time reflection symmetry. The conservation of the current is guaranteed by definition.

A numerical analysis of $T_{\mu\nu}$ in $d=3$, reveals that the stress tensor no longer has the form of an ideal fluid, i.e.
\begin{equation}
    T^{\mu\nu}(x) 
        =\frac{\sigma}{\pi n^{3}}\int_{0}^{2\pi} \hspace{-.25cm} d\theta \frac{\hat p^{\mu}\hat p^{\nu}}{\beta(x,\hat p)^{3}}=\text{Diag}\{\rho,P_x,P_{\perp}\}\;,\qquad \qquad t=0\;.
\end{equation}
where $P_x$ and $P_\perp$, with $P_x > P_\perp$, are the longitudinal and perpendicular pressures with respect to $x$ respectively, and $\rho$ is the energy density. For higher dimensions, the stress tensor is seen to remain diagonal, extended with identical $P_\perp$ on all dimensions orthogonal to $x$. The tracelesness of $T^{\mu\nu}$ imposes $\rho=P_x+(d-2)P_\perp$. 
This is often called an anisotropic fluid. Albeit anisotropic fluids can still be modeled by a combination of non-interacting perfect fluids, see e.g. \cite{Letelier1980}, we actually expect that for general (less symmetric) regions the tower of conserved currents cannot be interpreted as any particular family of fluid equations. The description of the system in terms of the tower of Lorentzian conserved currents is still valid for any region. 

\section{Summary and Conclusions}

Entanglement temperatures (ET) \cite{Arias:2016nip,arias2017anisotropic} are generalizations of Unruh temperatures associated with a point and a particular direction in a subsystem, and generated by entanglement with the rest of the global state. 
In \cite{Arias:2016nip,arias2017anisotropic}, the ET were defined from the leading term in the relative entropy between the state in a subsystem excited with a high energy and localized unitary and the state without the excitation, see eq. \eqref{hht}.
These entanglement temperatures follow from the solutions of a non-linear differential geometric problem involving complex vectors in Euclidean space. Analytic solutions of these equations, however, are in general out of reach at present, except for some highly symmetric examples. On the other hand, ET appear to be highly universal quantities.
In particular, these were shown to be the same for fermions and bosons, massive or massless \cite{arias2017anisotropic}.     

In this work, we have continued the study of ET. In particular, we find that there are an infinite number of ways of parametrizing the Euclidean solutions to the eikonal equations that define the local temperatures. We find no apparent preference for any representative of the family, since they all give place to the same ET. By considering deformations to free theories, we conclude that ET depend only on the conformal class determined by the metric, the region, and the UV conformal fixed point. All free UV fixed points share the same ET for a fixed geometry. For a general non-free UV fixed point, where a particle-like description of the UV physics is absent, we still lack a robust framework to compute ET. In this sense, it would be interesting to understand if there is a simple way to compute these temperatures for holographic theories. It is possible that ET arising from the eikonal equations could be completely universal, and valid even for interacting CFTs.

We have also studied the real-time propagation of the ET, which follows light-like geodesics and travels without interacting with each other at different times. In other words, the ET follow a ballistic propagation as a consequence of their ``eikonal'' nature. A temperature covariant vector $T^\mu$ can be defined as a light-like vector. Taking a high energy excitation with light-like momentum $p^\mu$, such that $p^\mu$ and $T^\mu$ are parallel,  the proportionality factor between these vectors is precisely the relative entropy of the excitation with respect to the vacuum. On the other hand, the Lorentzian problem by itself does not give enough information to uniquely determine the ET, it only determines their ballistic propagation. To find the initial condition for the problem we still rely on solutions to the Euclidean version of the problem. An example of local temperatures for a Vaidya metric was presented, in which the black hole temperature is recovered, and shown to change according to the considered subsystem.

Our main result in this work is the relationship between the ET and the limit of large modular temperature in the subsystem. This is described by the $n\to0$ limit of the Rényi entropies, Rényi-$0$ for short. This connection is nontrivial, in the sense that the original definition for ET are associated with highly energetic and localized excitations and not as quantities obtained from high-temperature systems. The Rényi-0 is obtained as an average of the ET in momenta and spacetime presented in \eqref{Rényi-zero-formula}. The Renyi entropies in the $n\to0$ limit are then computed as functions of the ET. This gives a universality for the Renyi-$0$ entropies. They are proportional to a geometric factor depending on the region and are computable from the ET. We reproduce several known results for this limit of the Renyi entropies by computing them in terms of ET.

Finally, using this high-temperature gas perspective, we review the derivation of the  free relativistic Boltzmann equation and show its natural connection with ET for theories with a free UV fixed point. The ballistic propagation of the ET in turn implies the conservation of an infinite tower of currents, defined as all possible weighted averages (moments in the sense of a distribution function) of the ET of the system, \eqref{Tmunu-Currents}. This is to say that the full tower of conserved currents contains in principle the same information as the ET. 
We explicitly built the tower of currents for a couple of examples. In the presence of a Killing vector in the geometry, the system of currents can be rewritten as the equations of a perfect fluid. Furthermore, given a single current in a scenario with a Killing vector, one can rebuild the full tower by symmetry arguments as shown in Sec. \ref{KillingCurrents}. This description fails for less symmetric systems.

It would be very interesting to recast the (highly non-linear) problem of finding the ET for a given region into a (hopefully simpler) problem of finding an adequate set of conserved currents in Euclidean spacetime. To this end, one could propose a Wick rotation of the Lorentzian tower of currents \eqref{Tmunu-Currents}, which should have vanishing divergence. A set of boundary conditions near the boundaries of the system and at infinity can also be extrapolated from the ones of the ET. However, much like a fluid, unless an adequate number of relations between currents or constraints are imposed (the constitutive equations) the system would not be well-posed on itself. We would like to explore this problem further in the future.

Another interesting perspective is the description of the large modular temperature limit for non-free CFTs. In this case, we do not expect a connection with ET because the large modular temperature produces high-energy excitations that are now interacting. Instead, we expect to have a description in terms of a  more general relativistic fluid. In this language, the free case corresponds to a free Boltzmann fluid. It would be interesting to understand the nature of these fluids for different CFTs.

\section*{Acknowledgments}
 This work was partially supported by CONICET, CNEA
and Universidad Nacional de Cuyo, Argentina. The work of H. C. 
 and C.A. was partially supported by an It From Qubit grant by the Simons foundation.

\appendix

\section{Canonical transformations of eikonal solutions for Rindler space}
\label{app1}

Consider the case of $d=3$ for simplicity. A simple solution for the eikonal equations for the case of half-space follows from rotation and translation invariance along the $z$ axis \cite{arias2017anisotropic}
\be
\alpha=\theta + i \left(k \, z+  \sqrt{1- k^2\,r^2}- \textrm{arctanh} \left(\sqrt{1- k^2\, r^2}\right)\right)\,,
\ee
where $r,\theta$ are polar coordinates on the plane $x^0,x^1$, and $x^3=z$. The real parameter $k\in(-\infty,\infty)$ gives the different solutions. Each solution has a compact range $r\le |k|^{-1}$ in the plane perpendicular to the $z$ axis. However, choosing different $k$ the ET for any point in the $x^0=0$ plane can be computed in any direction $\hat{B}$, with 
\be
\nabla \alpha=(A_0, i\, \vec{B})\;,\qquad\qquad x^0=0\;.
\ee
It follows the well-known result that $\beta=2\pi/A_0=2\pi r$, independently of direction. This is evident in this case because $\vec{A}=\nabla \,\textrm{Re}
\,\alpha=\nabla \theta$, that does not depend on $k$ that sets the direction of $\vec{B}$.

Now we make the canonical transformation induced by the function $f(k)=i\, L\, k$. By writing
$\tilde{\alpha}=\alpha+ i\,L\,k$, the equation $\partial \tilde{\alpha}/\partial k=0$ becomes
\bea
L+z=-\frac{\sqrt{1-k^2 r^2}}{k}\qquad \Rightarrow\qquad k=\pm\sqrt{r^2+(L+z)^2}\;,
\eea
where the $\pm$ sign is the opposite of the sign of $(L+z)$. 
Replacing this back in $\tilde{\alpha}$ we get the new solutions
\be
\tilde{\tilde{\alpha}}=\theta \pm i\, \textrm{arctanh}\left(\frac{L+z}{\sqrt{r^2+(L+z)^2}}\right)\,. 
\ee
This new set of solutions now has a range in the full space but gives place to the same entanglement temperatures. More explicitly the vectors $\vec{A}$ and $\vec{B}$ obtained from this new family are given by 
\begin{equation}\label{AB-L}
\vec{A}= \frac{\hat{\theta}}{r}\;,\qquad\qquad \vec{B}=\mp \frac{1}{r } \(\frac{(L+z)\hat{r}}{\sqrt{r^2+(L+z)^2}}- \frac{r\hat{z}}{\sqrt{r^2+(L+z)^2}}\)\;.
\end{equation}  

\bibliography{EE}{}
\bibliographystyle{utphys}

\end{document}